\documentclass[trackchanges]{aastex701} 

\usepackage{graphicx} 
\graphicspath{{figures/}}
\usepackage{subcaption} 
\usepackage{threeparttable}
\usepackage{enumitem}
\usepackage{amsmath}
\usepackage{makecell}
\usepackage{booktabs} 
\usepackage[
  separate-uncertainty = true,
  per-mode = symbol,
  uncertainty-mode = separate
]{siunitx} 
\DeclareSIUnit\year{yr}

\DeclareSIUnit\year{yr}
\DeclareSIUnit \parsec {pc}
\DeclareSIUnit \Jy {Jy}
\DeclareSIUnit \DM {\parsec\per\cubic\cm}
\DeclareSIUnit\dBi{dBi}
\DeclareSIUnit\dBm{dBm}
\DeclareSIUnit{\deg}{deg} 
\DeclareSIUnit \erg{erg}

\begin{document}

\title{Design and Commissioning of the BURSTT-256  Main Antenna Array \\  for the Fast Radio Burst Survey in the Local Universe }


\author[0000-0002-0060-7975]{Shih-Hao Wang}
\affiliation{Institute of Astronomy and Astrophysics, Academia Sinica,  Taipei 106319, Taiwan}
\affiliation{Leung Center for Cosmology and Particle Astrophysics, National Taiwan University, Taipei 106319, Taiwan}
\email{shwang@asiaa.sinica.edu.tw}

\author[0000-0002-8698-7277]{Kai-Yang Lin}
\affiliation{Institute of Astronomy and Astrophysics, Academia Sinica,  Taipei 106319, Taiwan}
\email{kylin@asiaa.sinica.edu.tw}

\author[0000-0002-5830-2226]{Chao-Te Li}
\affiliation{Institute of Astronomy and Astrophysics, Academia Sinica,  Taipei 106319, Taiwan}
\email{ctli@asiaa.sinica.edu.tw}

\author[0000-0002-5557-1439]{Chau-Ching Chiong}
\affiliation{Institute of Astronomy and Astrophysics, Academia Sinica,  Taipei 106319, Taiwan}
\email{ccchiong@asiaa.sinica.edu.tw}

\author{Jen-Chieh Cheng}
\affiliation{Institute of Astronomy and Astrophysics, Academia Sinica,  Taipei 106319, Taiwan}
\email{jccheng@asiaa.sinica.edu.tw}

\author{Chung-Cheng Chen}
\affiliation{Institute of Astronomy and Astrophysics, Academia Sinica,  Taipei 106319, Taiwan}
\email{mark@asiaa.sinica.edu.tw}

\author[0000-0003-4238-6469]{Hsien-Chun Tseng}
\affiliation{Institute of Astronomy and Astrophysics, Academia Sinica,  Taipei 106319, Taiwan}
\email{hctseng@asiaa.sinica.edu.tw}

\author{Shu-Hao Chang}
\affiliation{Institute of Astronomy and Astrophysics, Academia Sinica,  Taipei 106319, Taiwan}
\email{shchang@asiaa.sinica.edu.tw}

\author{Pai-An Chen}
\affiliation{Institute of Astronomy and Astrophysics, Academia Sinica,  Taipei 106319, Taiwan}
\email{paianchen@asiaa.sinica.edu.tw}

\author{Ta-Shun Wei}
\affiliation{Institute of Astronomy and Astrophysics, Academia Sinica,  Taipei 106319, Taiwan}
\email{tashun@asiaa.sinica.edu.tw}

\author{Chin-Ting Ho}
\affiliation{Institute of Astronomy and Astrophysics, Academia Sinica,  Taipei 106319, Taiwan}
\email{ctho@asiaa.sinica.edu.tw}

\author{Vigo Feng-Shun Wu}
\affiliation{Institute of Astronomy and Astrophysics, Academia Sinica,  Taipei 106319, Taiwan}
\email{fswu@asiaa.sinica.edu.tw}

\author[0000-0001-8783-6211]{Yu-Chieh Huang}
\affiliation{Institute of Astronomy and Astrophysics, Academia Sinica,  Taipei 106319, Taiwan}
\email{ydhuang@asiaa.sinica.edu.tw}

\author[0000-0003-0340-0651]{Sujin Eie}
\affiliation{Institute of Astronomy and Astrophysics, Academia Sinica,  Taipei 106319, Taiwan}
\affiliation{Max-Planck-Institut für Radioastronomie, Auf dem Hügel 69, D-53121, Germany}
\email{astrosujin@gmail.com}

\author[0000-0002-6495-8600]{Yao-Huan Tseng}
\affiliation{Institute of Astronomy and Astrophysics, Academia Sinica,  Taipei 106319, Taiwan}
\email{yhtseng@asiaa.sinica.edu.tw}

\author[0000-0003-2155-9578]{Ue-Li Pen}
\affiliation{Institute of Astronomy and Astrophysics, Academia Sinica,  Taipei 106319, Taiwan}
\affiliation{Canadian Institute for Theoretical Astrophysics, 60 St. George Street, Toronto, ON M5S 3H8, Canada}
\affiliation{Canadian Institute for Advanced Research, MaRS Centre, West Tower, 661 University Avenue, Suite 505}
\affiliation{Dunlap Institute for Astronomy and Astrophysics, University of Toronto, 50 St George Street, Toronto, ON M5S 3H4, Canada}
\affiliation{Perimeter Institute of Theoretical Physics, 31 Caroline Street North, Waterloo, ON N2L 2Y5, Canada}
\email{pen@asiaa.sinica.edu.tw}

\author[0000-0003-4058-3303]{Chih-Yi Wen}
\affiliation{Institute of Astronomy and Astrophysics, Academia Sinica,  Taipei 106319, Taiwan}
\email{wency@asiaa.sinica.edu.tw}

\author[0000-0002-9893-2433]{Homin Jiang}
\affiliation{Institute of Astronomy and Astrophysics, Academia Sinica,  Taipei 106319, Taiwan}
\email{homin@asiaa.sinica.edu.tw}

\author[0000-0003-4708-5964]{Andrew Wang}
\affiliation{Institute of Astronomy and Astrophysics, Academia Sinica,  Taipei 106319, Taiwan}
\email{andrew@asiaa.sinica.edu.tw}

\author[0009-0005-4098-8396]{Bishnu Kumar Sharma}
\affiliation{Institute of Astronomy and Astrophysics, Academia Sinica,  Taipei 106319, Taiwan}
\email{bksharma@asiaa.sinica.edu.tw}

\author[]{Tzu-Hong Chen Lin}
\affiliation{Institute of Astronomy and Astrophysics, Academia Sinica,  Taipei 106319, Taiwan}
\email{thchenlin@asiaa.sinica.edu.tw}

\author[]{Yuan-Wei Hu}
\affiliation{Taiwan Forestry Research Institute, No. 53, Nanhai Road, Taipei City 10066, Taiwan, Taiwan}
\email{why64@tfri.gov.tw}

\author[]{Shih-Wen Chung}
\affiliation{Taiwan Forestry Research Institute, No. 53, Nanhai Road, Taipei City 10066, Taiwan, Taiwan}
\email{biflora@tfri.gov.tw}

\author[]{Chen-Hui Peng}
\affiliation{Taiwan Forestry Research Institute, No. 53, Nanhai Road, Taipei City 10066, Taiwan, Taiwan}
\email{elsa@tfri.gov.tw}



\author[0000-0001-6573-3318]{Ming-Tang Chen}
\affiliation{Institute of Astronomy and Astrophysics, Academia Sinica,  Taipei 106319, Taiwan}
\affiliation{Institute of Astronomy and Astrophysics, Academia Sinica, 645 North Aohoku Place, Hilo, HI 96720 USA}
\email{mtchen@asiaa.sinica.edu.tw}

\author[0000-0001-5251-7210]{Pisin Chen}
\affiliation{Leung Center for Cosmology and Particle Astrophysics, National Taiwan University, Taipei 106319, Taiwan}
\affiliation{Department of Physics, National Taiwan University, Taipei 106319, Taiwan}
\email{pisinchen@phys.ntu.edu.tw}

\author[0000-0001-9673-3134]{Jiwoo Nam}
\affiliation{Leung Center for Cosmology and Particle Astrophysics, National Taiwan University, Taipei 106319, Taiwan}
\affiliation{Department of Physics, National Taiwan University, Taipei 106319, Taiwan}
\email{jwnam@phys.ntu.edu.tw}

\author[0000-0002-6821-8669]{Tomotsugu Goto}
\affiliation{Institute of Astronomy, National Tsing Hua University, No. 101, Section 2, Kuang-Fu Road, Hsinchu City 30013, Taiwan}
\email{tomo@phys.nthu.edu.tw}

\author[0000-0001-7228-1428]{Tetsuya Hashimoto}
\affiliation{Department of Physics, National Chung Hsing University, No. 145, Xingda Rd., South Dist., Taichung, 40227, Taiwan}
\email{tetsuya@phys.nchu.edu.tw}

\author[0000-0001-8551-2002]{Chin-Ping Hu}
\affiliation{Department of Physics, National Changhua University of Education, Changhua, 50007, Taiwan}
\email{cphu0821@cc.ncue.edu.tw}

\author[0000-0002-0769-8283]{Yuan-Chang Hou}
\affiliation{Department of Electrical Engineering, National Ilan University, Yilan 260007, Taiwan}
\email{ychou@niu.edu.tw}

\author[0000-0001-7888-3470]{Daniel Baker}
\affiliation{Institute of Astronomy and Astrophysics, Academia Sinica,  Taipei 106319, Taiwan}
\email{dbaker@asiaa.sinica.edu.tw}

\author[]{Michael Chiang}
\affiliation{Institute of Astronomy and Astrophysics, Academia Sinica,  Taipei 106319, Taiwan}
\email{ccchiang@asiaa.sinica.edu.tw}

\author[0000-0002-2338-4583]{Somnath Dutta}
\affiliation{Institute of Astronomy and Astrophysics, Academia Sinica, Taipei 106319, Taiwan}
\email{duttasomnath9@gmail.com}

\author[0000-0001-9461-1141]{Sun-Kun King}
\affiliation{Institute of Astronomy and Astrophysics, Academia Sinica,  Taipei 106319, Taiwan}
\email{skking@ms10.hinet.net}

\author[0009-0003-5746-1510]{Hiroto Masaoka}
\affiliation{Department of Astronomy, School of Science, The University of Tokyo, 7-3-1 Hongo, Bunkyo-ku, Tokyo 113-0033, Japan}
\affiliation{Mizusawa VLBI Observatory, National Astronomical Observatory of Japan, 1-2-2 Mizusawa-Hoshigaoka, Oshu, Iwate 023-0861, Japan}
\email{hmasaoka@g.ecc.u-tokyo.ac.jp}

\author[0000-0002-0563-4497]{Chung-Yun Kuo}
\affiliation{Leung Center for Cosmology and Particle Astrophysics, National Taiwan University, Taipei 106319, Taiwan}
\affiliation{Department of Physics, National Taiwan University, Taipei 106319, Taiwan}
\email{chungyunkuo@ntu.edu.tw}

\author[0009-0001-4371-6573]{Che-Wei Pai}
\affiliation{Leung Center for Cosmology and Particle Astrophysics, National Taiwan University, Taipei 106319, Taiwan}
\affiliation{Department of Physics, National Taiwan University, Taipei 106319, Taiwan}
\email{r13244002@ntu.edu.tw}

\author[]{Decmend Fang-Jie Ling}
\affiliation{Institute of Astrohysics, National Taiwan University, Taipei 106319, Taiwan}
\email{decmend01@gmail.com}

\begin{abstract}

The Bustling Universe Radio Survey Telescope in Taiwan (BURSTT) is a radio interferometer array designed for surveying fast radio bursts (FRBs) in the nearby Universe with a wide instantaneous field of view of $\approx$\SI{6000}{\deg\squared} and precisely localizing them with sub-arcsecond angular resolution.
The first phase of BURSTT, BURSTT-256, consists of a main antenna array comprising 256 log-periodic dipole antennas (LPDAs) observing at \SIrange{400}{800}{\MHz} for real-time beamforming and FRB search, currently with 256 beams covering  $\approx$\SI{3200}{\deg\squared} field of view (FoV), together with several smaller outrigger stations with 16-64 antennas and separated by $\mathcal{O}(10^2\text{--}10^3)$\si{\km} baselines for joint observations to localize FRBs using very-long baseline interferometry (VLBI).
This work presents the design, deployment, and commissioning of the BURSTT-256 main array, including  site selection, radio-frequency (RF) background characterization, receiver antenna array and RF front-end, as well as real-time, factorized digital beamforming backend that currently generates 256 continuous monitoring beams.
The system's on-sky performance, sensitivity, and beamforming fidelity have been calibrated through observations of bright celestial calibrators.
The antenna array in the initial phase using commercial LPDAs has a system equivalent flux density of $\approx$\SI{11}{\kilo\Jy}, and is expected to achieve $\approx$\SI{4}{\kilo\Jy} using custom-design LPDAs after optimization in late 2026.
BURSTT-256 serves as a critical pathfinder for future expansion of the array to 2048 elements and beyond, operating at \SIrange{300}{700}{\MHz} with $\approx$\SI{6000}{\deg\squared} FoV. It establishes the framework required to discover and localize nearby FRBs with sub-arcsecond precision. By substantially increasing the number of well-localized nearby FRBs, BURSTT will facilitate multi-wavelength and multi-messenger counterpart identification, helping to unveil the elusive origin of FRBs.

\end{abstract}

\keywords{\uat{Fast radio burst}{2008}, \uat{Astronomical instrumentation}{799}, \uat{Radio telescopes}{1360}, \uat{Radio astronomy}{1338}}

\setcounter{footnote}{0} 

\section{Introduction}

\subsection{Fast Radio Bursts}
\label{sec:FRB}


Fast radio bursts (FRBs) are millisecond-duration radio transients of extragalactic origin, detected from \SI{110}{\MHz} \citep{Pleunis2021} up to \SI{8}{\GHz} \citep{Gajjar2018}, with immense energy output of isotropic equivalent energies of $10^{28}\text{--}10^{36}$ \si{\erg}) \citep{Zhang2023}. 
As these coherent radio pulses traverse the intervening ionized astrophysical plasma, they undergo frequency-dependent dispersion which induces a characteristic quadratic arrival-time delay, $\Delta t \propto DM \cdot \nu^{-2}$ \citep{Zhang2023}. The magnitude of this delay is quantified by the dispersion measure ($\mathrm{DM} \equiv \int n_{e}dl$), representing the integrated column density of free electrons along the line of sight.
FRBs typically exhibit DMs of $\mathcal{O}(10^2\text{--}10^3)$  \si{\DM}, with substantial excesses over the Galactic plasma contribution that confirm their extragalactic nature \citep{Cordes2019}.
Their coherent pulses propagating over cosmological distances, currently confirmed up to redshift $z\approx 2$ (\citealt{Caleb2025}), can potentially serve as a new cosmological probe, e.g., mapping the baryon distribution of the Universe \citep{Lee2022,Hsu2025,Hussaini2025}. For comprehensive reviews of the field, readers are referred to \cite{Petroff2019,Cordes2019,Petroff2022,Zhang2023,Zhang2024}.

Since the first discovery in 2007 (\citealt{Lorimer2007,Lorimer2024}), more than 4000 FRBs have been discovered to date, with the vast majority (3641 unique sources) contributed by the Canadian Hydrogen Intensity Mapping Experiment (CHIME) FRB project (\citealt{CHIME2026}). 
The estimated all-sky FRB rate is $\sim 1000$ per day for fluence above \SI{5}{\Jy \milli\second} (\citealt{CHIME1stCatalog}).
Despite the large number of samples, the precise physical origins of FRBs remain elusive. For example, a key phenomenological division exists within the population: while most FRBs appear as one-off events, about dozens of FRB sources emit multiple pulses and thus are called as ``repeaters'', suggesting their non-catastrophic origin. Repeaters are found to exhibit emission properties distinct from non-repeaters, such as wider temporal durations and narrower spectral bandwidths \citep{CHIME2021}. This dichotomy prompts debate over whether repeaters and non-repeaters originate from distinct astrophysical populations or multiple progenitor formation channels, e.g., \cite{Hashimoto2020,Yamasaki2024}.
It could be possible that repeating FRBs appear as one-off events due to the lack of a comprehensive FRB survey so far.
To address this issue, a wide field-of-view instrument is important because it allows long-term continuous monitoring for each FRB source.

%
Local environments of FRB sources provide insightful implications for constraining the progenitor types. 
To scrutinize local environments of FRBs and fully characterize FRB's intrinsic properties, localizing FRBs with arcsecond angular resolution or beyond is necessary to identify their host galaxies and even to the region within, and  to facilitate coincident multi-wavelength and multi-messenger follow-up observations.
To date, only more than a hundred of FRBs have been precisely localized to their host galaxies \citep{Lorimer2024,DSA2024b,CHIME-KKO2025}, revealing a highly diverse range of environments.
%
However, most of localized FRBs are distant, with only a few exceptions within $\sim$\SI{100}{\mega\parsec} distance, e.g., one Galactic FRB 200428 which is associated with the Galactic magnetar SGR 1935+2154  \citep{CHIME2020,STARE2020} with a coincident multi-wavelength (X-ray) counterpart observed \citep{INTEGRAL2020,Insight2021,Konus2021}; and FRB 20200120E localized in a globular cluster in the nearby M81 galaxy (\citealt{CHIME2022}); repeating FRB 20181030A hosted by NGC 3252 at $\sim$\SI{20}{\mega\parsec} distance \citep{CHIME2021b}; and FRB 20231229A at $\sim$\SI{90}{\mega\parsec} distance \citep{CHIME-KKO2025}.

The lack of well-localized FRBs in the local Universe is mainly because most, if not all, of existing telescopes discovering the FRBs are not specifically designed for FRB surveys.
Traditional large-aperture single-dish telescopes --- such as Parkes, Green Bank, and FAST telescopes --- offer high sensitivity to detect fainter bursts at farther distances to probe larger cosmological volumes. However, they alone do not have adequate angular resolution to localize FRBs to their host galaxies, and their small field of view (FoV) limits their ability to monitor the sky with high cadence or for a long time, and consequently, repeaters may get misclassified as one-off events.
Telescopes consisting of arrays of dish antennas, either ones dedicated to FRB survey, e.g., Deep Synoptic Array-110 (DSA-110; \citealt{DSA2024a,DSA2024b}), or those in commensal mode, e.g., CRAFT at Australian SKA Pathfinder (ASKAP; \citealt{CRAFT2025}), MeerTRAP at MeerKAT (\citealt{MeerTRAP2022}), and SPOTLIGHT at Giant Metrewave Radio Telescope (GMRT; \citealt{Panda2024}), improve the localization capability with the Very Long Baseline Interferometry (VLBI) technique, but their FoV is still limited to $\mathcal{O}(10)$  \si{\deg\squared}.
CHIME/FRB \citep{CHIME2018} has a wider FoV, $\sim 110^\circ\times 2^\circ$, and recently with outrigger stations  \citep{CHIME2025a} to improve its localization resolution to $\sim$\ang{;;0.05}. However, the FoV is still narrow for long-term monitoring or for covering a substantial amount of the full sky.
All of these highlight the necessity of a dedicated telescope which has both wide FoV and localization capabilities.


\subsection{BURSTT}

Motivated by the need to identify FRB progenitors, elucidate their emission mechanisms, and characterize their local host environments, the Bustling Universe Radio Survey Telescope in Taiwan (BURSTT) project was initiated \citep{BURSTT2022}, a radio telescope designed to survey and localize FRBs in the nearby Universe. 
BURSTT features a wide instantaneous field of view of $\approx$\SI{6000}{\deg\squared}, which is roughly 27 times larger than that of CHIME telescope, and precise FRB localization with sub-arcsecond angular resolution, for not only identifying host galaxies but also the source regions.
To achieve wide FoV, the system employs real-time beamforming antenna arrays composed of lower-gain elements, which support a fixed \SI{400}{\MHz} instantaneous processing bandwidth that can be positioned anywhere within the \SIrange{300}{800}{MHz} frequency range. This capability allows for continuous monitoring of a vast portion of the sky, and thus is able to monitor known FRBs for long term with high cadence.
The precise FRB localization is achieved by building multiple beamforming antenna arrays in Taiwan and throughout the Asia-Pacific region, with a main station and multiple outrigger stations distributed over baselines greater than \SI{100}{\km}.

The expected detection range of BURSTT is at the median redshift $z \sim 0.04$, or $\sim \SI{170}{\mega\parsec}$ distance, with an estimated FRB event rate between several and 100 events per year \citep{BURSTT2022,BURSTT_overview}, which will significantly increase the number of well-localized FRBs.
%
Precise FRB localization will facilitate multi-wavelength and multi-messenger follow-up or coincident observations.
This distance range is strategically chosen to overlap with the sensitivity horizon of current LIGO-Virgo-KAGRA Gravitational Wave (GW) observatories\footnote{\href{https://observing.docs.ligo.org/plan/}{https://observing.docs.ligo.org/plan/}}, thereby enhancing the probability of detecting coincident multi-messenger events (e.g., \citealt{Zhang2024}). Furthermore, the telescope's location in the Asia-Pacific region fills the longitudinal gap in global sky coverage, complementing existing observatories like CHIME/FRB.

This paper focuses on the deployment and commissioning of the BURSTT-256 main array during the first phase of the project in 2024--2025. Construction began in 2022, utilizing 256 commercial LPDAs with single polarization to form the primary receiving array with an effective aperture of $\sim$\SI{20}{\square\m}. The system employs a real-time factorized beamforming approach to manage the data stream and to generate tiled beams to cover the FoV.
%
Real-time pulse searching operations of BURSTT commenced in April 2025. The system has successfully validated its performance through the regular detection of single pulses from the pulsar B0329+54 and Giant Radio Pulses (GRPs) from the Crab pulsar (B0531+21), with the detailed architecture of the back-end system and the pulse-searching algorithms described in \cite{BURSTT_BACKEND}. 
The first FRB detection of BURSTT and the overview of all other stations will be presented in the upcoming paper ( \citealt{BURSTT_overview},in preparation).
The initial result on the VLBI will be presented in another paper (\citealt{BURSTT_VLBI}, in preparation).

This paper is organized as follows: Section \ref{sec:burstt256} describes the design of BURSTT-256 main array: from the site selection and RF background characterization (Sec.~\ref{sec:site}); the array configuration (Sec.~\ref{sec:array}) and the characteristics of two types of constituent antennas adopted at the main site --- the commercial LPDA for the early stage and a custom-deign LPDA which will be used for the next phase (Sec.~\ref{sec:Ant}). The analog receiver signal chain is introduced in Sec.~\ref{sec:FEM}, and then the architecture of the back-end system is briefly described in Sec.~\ref{sec:backend}. Section \ref{sec:cal} presents the calibration methods adopted by BURSTT, and initial  results of the commissioning and on-sky performance verification, followed by a summary and prospect in Section \ref{sec:conclusion}.

\section{Receiving System at BURSTT-256 Main Station}
\label{sec:burstt256}

\subsection{Overview}

The goal of the BURSTT receiving system is to monitor and record RF signals with \SI{400}{\MHz} instantaneous bandwidth within the 300–800 MHz frequency range. Since most FRBs are not repeating and with disperse signals that typically last for $\mathcal{O}(0.1\text{--}10)$ seconds, a system that can monitor the full sky all the time with high angular resolution is desired. A large array of dipole antennas is thus an ideal receiver choice for this purpose.

Every BURSTT station is a multi-beam phased array performing beamforming in real time. The main array with the most antennas and thus the highest sensitivity is responsible for real-time FRB detection, with signal processing algorithms provided by the back-end software (see \citealp{BURSTT_BACKEND} for details).
The characteristics of BURSTT main station is summarized in Table \ref{tab:summary}.
Once an FRB is detected by the main array, a trigger signal is sent to all outrigger stations to record the buffered baseband voltage data, currently with a duration of $\approx 16$ seconds, for further source localization analysis with the VLBI technique \citep{BURSTT_VLBI}, e.g., an initial test with the outrigger in Japan described in \cite{BURSTT_OGA16} (in preparation).

\begin{figure}[ht!]
    \centering
    \includegraphics[width=0.98\textwidth]{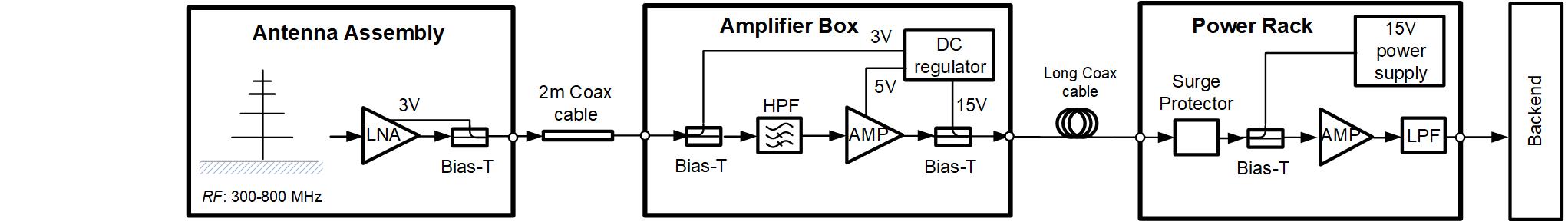}
    \caption{Block diagram of the BURSTT receiver front-end with ASIAA LPDA. For commercial LPDA, the LNA is housed in the amplifier box.
    \label{fig:front-end}}
\end{figure}

The block diagram of the BURSTT-256 receiver antenna and the RF front-end is shown in Figure~\ref{fig:front-end}. It consists of three parts: (1) the antenna assembly with a low-noise amplifier (LNA), (2) the front-end module (FEM) with a second-stage amplifier in a box near the antenna assembly, and (3) the third-stage amplifier with surge protector  housed in the outdoor rack enclosure. There are spaces for optional RF filters in the amplifier box and the outdoor rack enclosure. The system is in a single-ended configuration using coaxial connection.

\begin{table}[htbp]
  \centering
  \caption{Key system parameters of BURSTT-256 main station. Values outside parentheses correspond to the target full-array configuration with custom-design LPDAs expected in late 2026, whereas values in parentheses denote the initial phase configuration with commercial LPDAs in 2024--2025 for which results are presented in this work.}
  \label{tab:summary}
  \begin{tabular}{lc} 
    \toprule
    \textbf{Parameter} & \textbf{Value} \\
    \midrule
    Location  &  Fushan Botanical Garden, Yilan County, Taiwan \\
    WGS84 coordinates &  \ang{24;45;23.377} N,	\ang{121;34;53.921} E, \SI{642}{\m} \\
    Number of elements           & 256 \\
    Array configuration          & rectangular, $16 \times 16$ \\
    Array azimuthal offset   &  \ang{3.0} \\
    Element spacing ($d_X$, $d_Y$)        & \SI{1.0}{\m} along E-W, \SI{0.5}{\m} along N-S  \\
    Antenna element    &  ASIAA LPDA (commercial LPDA) \\
    Antenna gain ($G_{\mathrm{ant}}$)   & $\approx$\SI{8}{\dBi} ($\approx$\SI{6}{\dBi}) \\
    Polarization    &  single, linear along E-W axis of array \\

    \addlinespace
    Frequency range           & \SIrange{300}{700}{\MHz} (\SIrange{400}{800}{\MHz}) \\
    Number of frequency channels    &   1024 \\
    Frequency resolution ($\Delta\nu$)    & \SI{0.390625}{\MHz} \\
    Time resolution ($\tau$)     &   \SI{2.56}{\micro\s} \\
    ADC resolution   & 14 bits \\
    Quantization dynamic range   & 4 bits \\
    Ring buffer depth &     $\approx$\SI{15}{\s} \\

    \addlinespace
    Number of synthesized beams  &   512 (256) \\
    Beam width  &   $\approx\ang{2.5} \times \ang{5}$ \\
    Field of view   & $\approx$\SI{6000}{\deg\squared} ($\approx$\SI{3200}{\deg\squared})  \\
    Beamformed SEFD at \SI{400}{\MHz}    &  $\approx$\SI{4}{\kilo\Jy} ($\approx$\SI{11}{\kilo\Jy}) \\
    FRB search & since May 2025 \\
         
    \bottomrule
  \end{tabular}
\end{table}

\subsection{Site Selection and Development}
\label{sec:site}

\begin{figure}[htb!]
    \centering
    \includegraphics[width=0.9\linewidth]{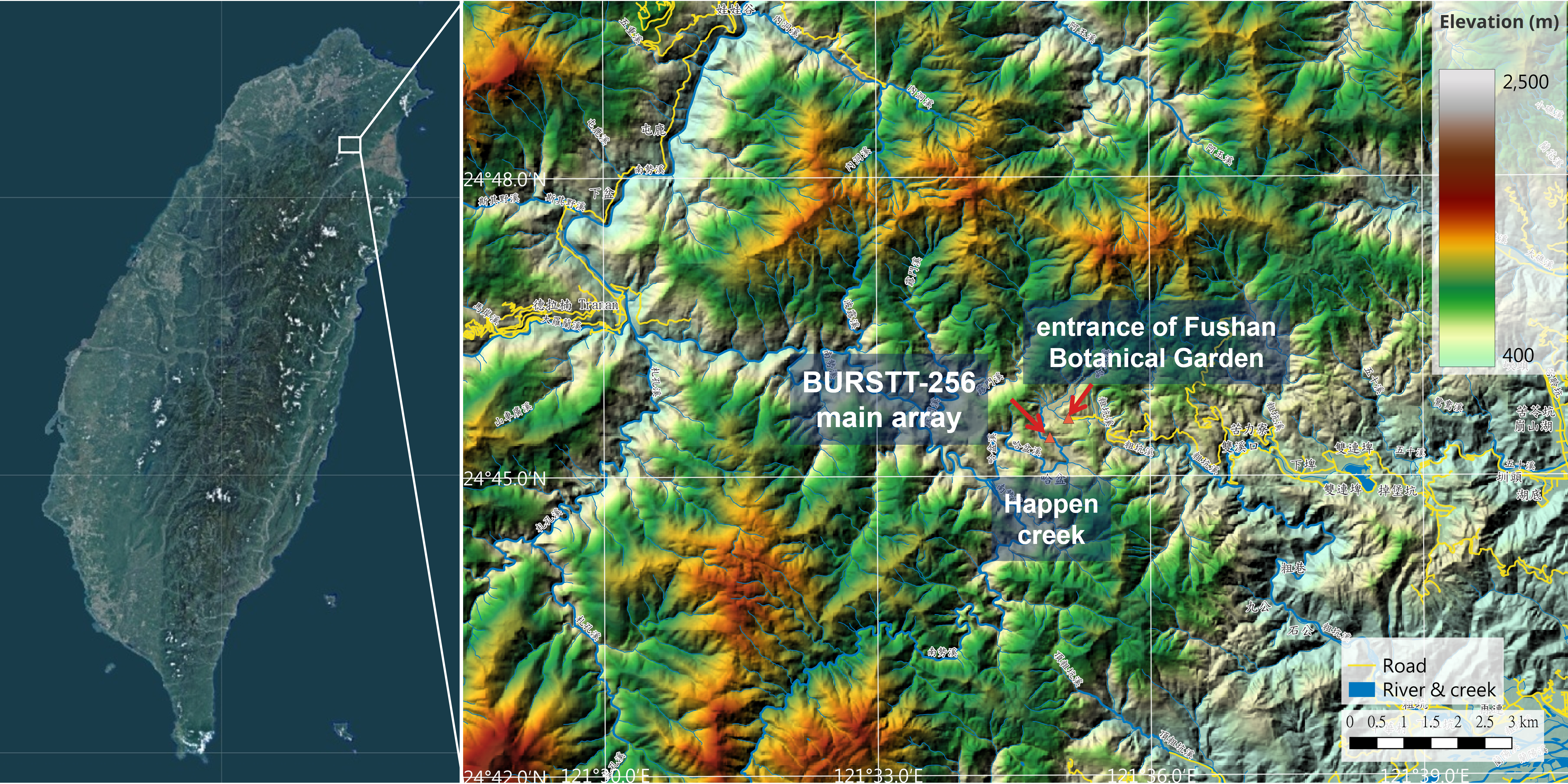}
    \caption{Location of the BURSTT-256 main site at Fushan Botanical Garden in Taiwan, encompassing conservation area surrounded by mountains (west to the entrance). The antenna array is $\sim$\SI{500}{\m} away from the entrance where the back-end servers are stored. Left map from \href{https://map.tgos.tw/TGOSCloudMap}{TGOS map}; the topographic map was produced with \textsc{QGIS} \citep{QGIS}, utilizing the open-access \SI{20}{\m} DEM from the Ministry of the Interior, Taiwan \citep{MOI_DEM_20m}, and vector features from OpenStreetMap \citep{OSM}.
    \label{fig:FushanMap}}
\end{figure}

During the initial phase of the project, site surveys were conducted in eastern Taiwan, primarily in Yilan and Hualien counties. These regions exhibit significantly less human activity than western Taiwan, resulting in a cleaner radio environment with weaker radio frequency interference (RFI). Ultimately, the Fushan Botanical Garden in Yilan  (\ang{24;45;23.377} N,	\ang{121;34;53.921} E) was chosen as the BURSTT's main site, as shown in Figure~\ref{fig:FushanMap}.
The garden encompasses a soil and water conservation area of approximately \SI{10}{\square\km} along the Hapen Creek; crucially, this designation restricts public access while permitting scientific research. 
Within this protected area, a small alluvial plain adjacent to the creek was selected to accommodate the main 256-element antenna array. 

\begin{figure}
    \centering
    \includegraphics[width=0.9\textwidth]{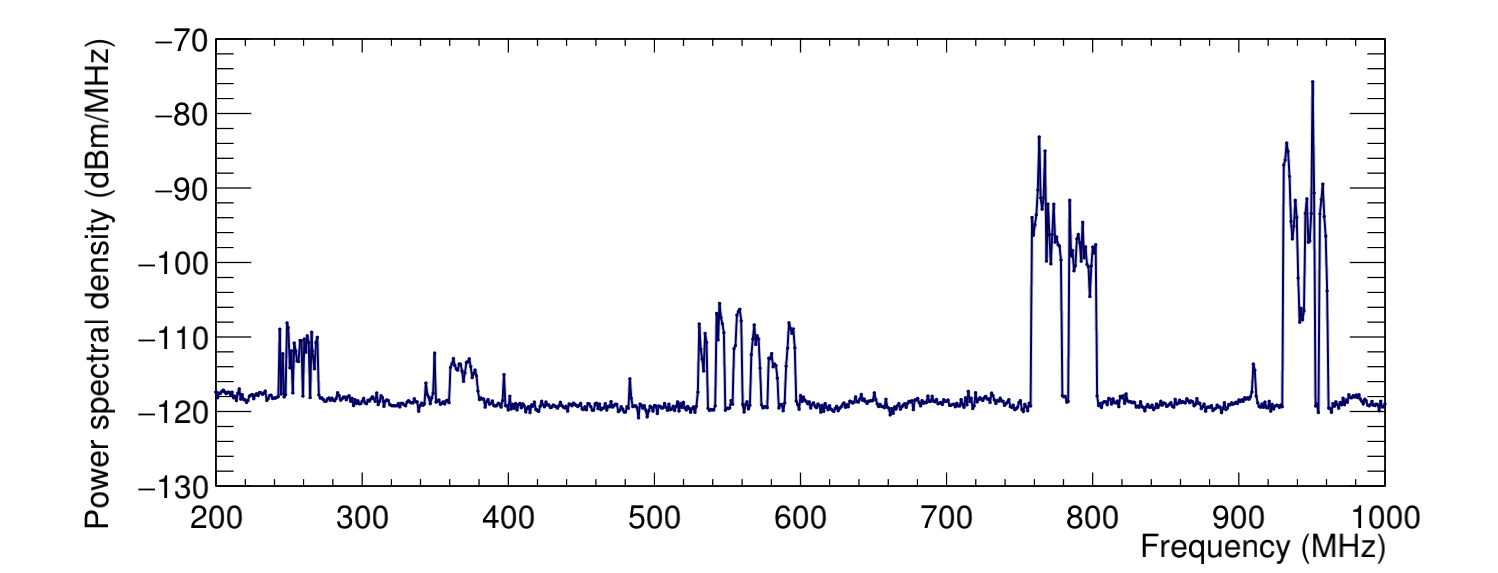}
	\caption{
		RF background spectrum measured at the valley of Hapen Creek of Fushan Botanical Garden on Dec 9, 2021, where BURSTT's main site was later deployed.
        The spectrum was measured with a spectrum analyzer connecting to an upward-pointing LPDA followed by a low-noise amplifier module, whose gain has been calibrated.
        The major RFI is from satellite communication at \SIrange{240}{270}{\MHz} and \SIrange{360}{380}{\MHz}, digital TV at \SIrange{520}{600}{\MHz}, and 4G-LTE mobile at \SIrange{750}{800}{\MHz} and \SIrange{930}{960}{\MHz}. 
	}
\label{fig:RFsurvey}
\end{figure}

The topography of the surrounding valley provides a natural shield against pervasive RFI sources, particularly digital TV (DTV) at \SIrange{520}{600}{\MHz} and 4G-LTE mobile communication, primarily the downlink at \SIrange{750}{800}{\MHz} and \SIrange{930}{960}{\MHz}\footnote{ The Republic of China Radio Frequency Allocation Table: \url{https://data.gov.tw/en/datasets/173237}. }. 
Although a 4G-LTE repeater tower is located approximately \SI{400}{\m} from the array at the garden, it is located on the opposite side of a hill, providing further isolation.
The RF background spectrum measured during the site survey is presented in Figure~\ref{fig:RFsurvey}. 
The result indicates that DTV signal levels are moderate and thus no dedicated hardware filtering is required for operation. However, a low-pass filter is required in the front-end module to prevent 4G-LTE signal from saturating the RF amplifiers at later stages. 
The RFI at \SIrange{240}{270}{\MHz} and \SIrange{360}{380}{\MHz} is ubiquitous in all surveyed sites and thus is likely to originate from the downlink signal of the military satellite communication (Satcom) by the Mobile User Objective System (MUOS; \citealp{Oetting2011}). 
Aside from these RFI bands, the Fushan site provides a wide usable bandwidth of $\sim$\SI{330}{\MHz} for FRB detection.

\begin{figure}[htb!]
    \centering
    \begin{subfigure}{.48\textwidth}
        \includegraphics[width=\textwidth ]{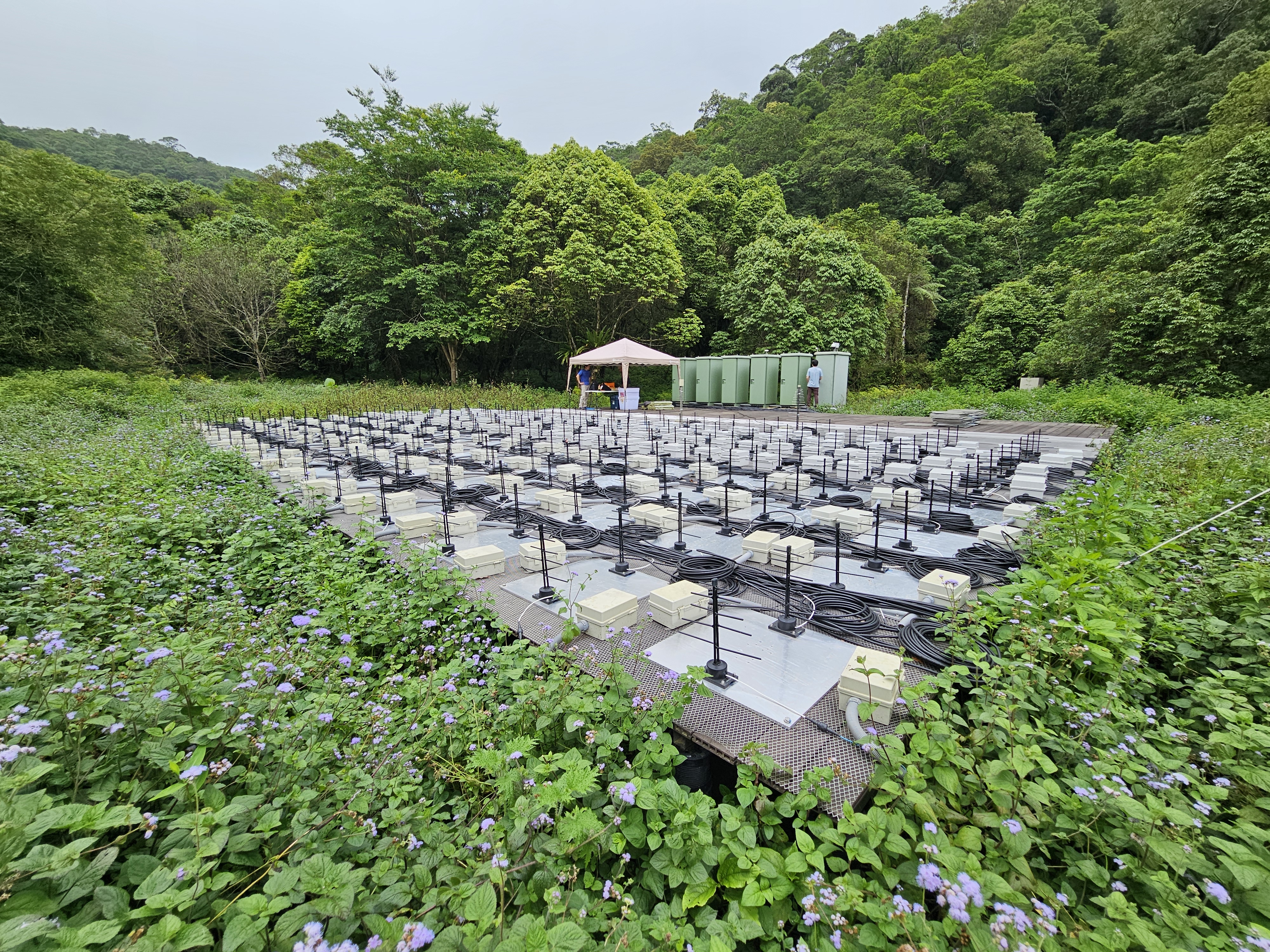}
    \end{subfigure}
     \begin{subfigure}{.48\textwidth}
        \includegraphics[width=\textwidth ]{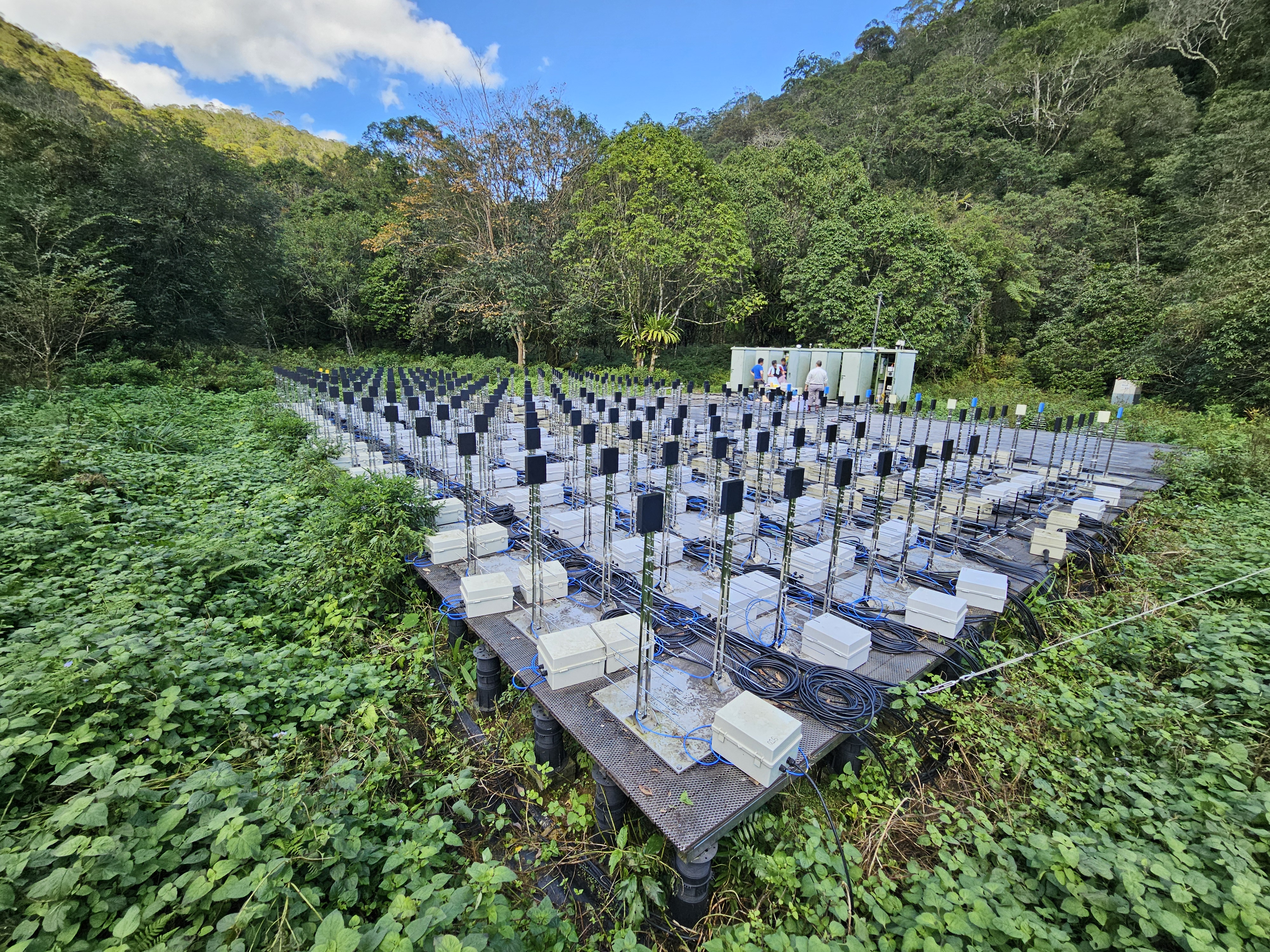}
    \end{subfigure}
    \caption{
        Left: initial phase of BURSTT-256 antenna array at Fushan with commercial LPDAs, deployed in May, 2024.
        Right: the antennas have been being progressively replaced by custom-designed LPDAs with higher gain and wider frequency coverage.
        The photos were taken at the southeast side of the array. The outdoor cabinets housing RFSoC boards and other electronics are in the background.
    }
    \label{fig:Ant_Layout}
\end{figure}

The construction of the site commenced in 2022, beginning with land preparation, followed by the installation of an elevated \SI{20}{\m} $\times$ \SI{16}{\m} wood-plastic composite platform to prevent potential flood damage. A layer of metal mesh was laid across the platform to serve as the ground shield to reduce thermal noise pickup from the back lobe of antennas. Power cables and optical fibers were routed from the garden entrance to the platform, where several outdoor cabinets were installed to house the system electronics. 
Currently, signal digitization is performed on site using Xilinx ZCU216 field programmable gate array (FPGA) based, RF System-on-Chip (RFSoC) boards located inside the cabinets. The digitized signals are transferred to the servers housed near the entrance of the garden, $\approx$\SI{500}{\m} away from the site and separated by hills, for further processing and real-time FRB search (Sec.~\ref{sec:backend}).  However, the back-end system on site generates a perceivable level of RFI. To mitigate this, the future phase of the project will relocate the digital electronics away from the site using RF over Fiber (RFoF) technology. The full array of 256 antennas and their associated front-end electronics were successfully installed in May 2024 (see Figure~\ref{fig:Ant_Layout}). 

\subsection{Receiving Antennas}
\label{sec:antenna}


\subsubsection{Array Configuration}
\label{sec:array}

The LPDA is chosen as the receiving element for the BURSTT array because of its favorable wideband frequency coverage. 
The main array consists of 256 antennas in a $16\times16$ rectangular grid.
The elements are spaced by \SI{1}{\m} along the East-West (E-W) axis and \SI{0.5}{\m} along the North-South (N-S) axis. All LPDAs share a common orientation, with the boresight pointing toward the zenith and the dipole elements aligned parallel to the E-W axis.
The 16 antennas within each E-W row are routed to a single RFSoC board located in the nearby outdoor cabinet for signal digitization and processing (Sec.~\ref{sec:backend}).
Post-deployment verification of the orientation of the entire array, performed using photogrammetry and astronomical source calibration (see Sec.~\ref{sec:BFCal}), revealed a systematic azimuthal offset of \ang{3}, i.e., the E-W axis is slightly in WNW-ESE direction.

\subsubsection{Antenna Elements}
\label{sec:Ant}
Two distinct LPDA designs have been employed during the development of the BURSTT-256 array: a commercial off-the-shelf model (Mobile Mark Y42400WB) and a custom high-performance model developed in-house at the Institute of Astronomy and Astrophysics, Academia Sinica (hereafter, the ASIAA LPDA).
To expedite the initial system deployment and validation, the commercial LPDA was utilized. The first full array of 256 antennas was deployed in May 2024.
Unless otherwise specified, the results presented in this paper refer to this initial configuration using commercial antennas.

\begin{figure}[htb!]
    \centering
        \includegraphics[width=0.75\textwidth ]{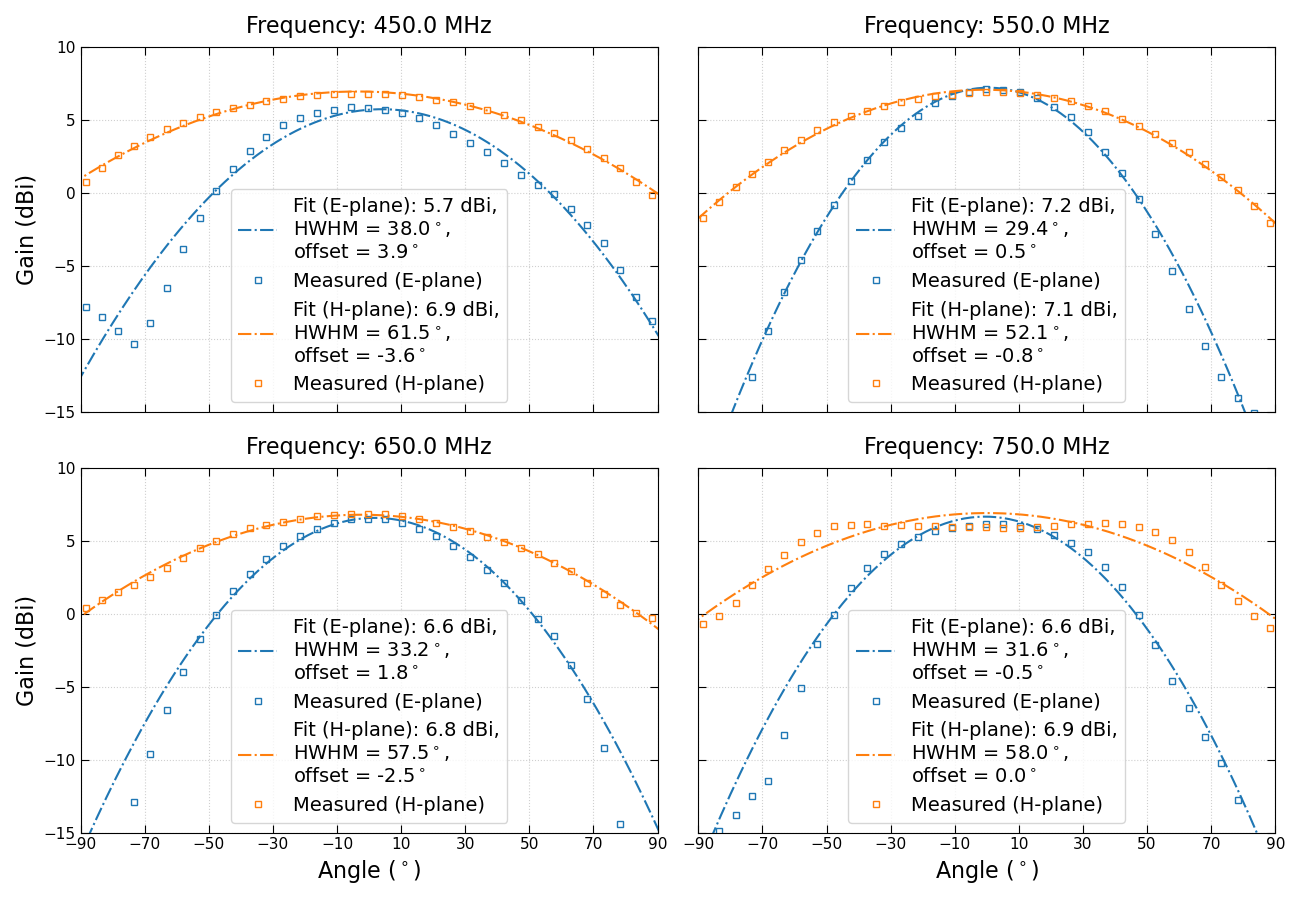}
    \caption{
        The measured E-plane (orange squares) and H-plane (blue squares) radiation pattern of commercial LPDA at 450, 550, 650, 750 \si{MHz} calibrated in RF anechoic chamber.
        The dashed curves of the same color are the corresponding Gaussian fit profile around the main lobe.
    }

    \label{fig:RadPattern-MM}
\end{figure}

\begin{figure}[htb!]
    \centering

        \includegraphics[width=0.75\textwidth ]{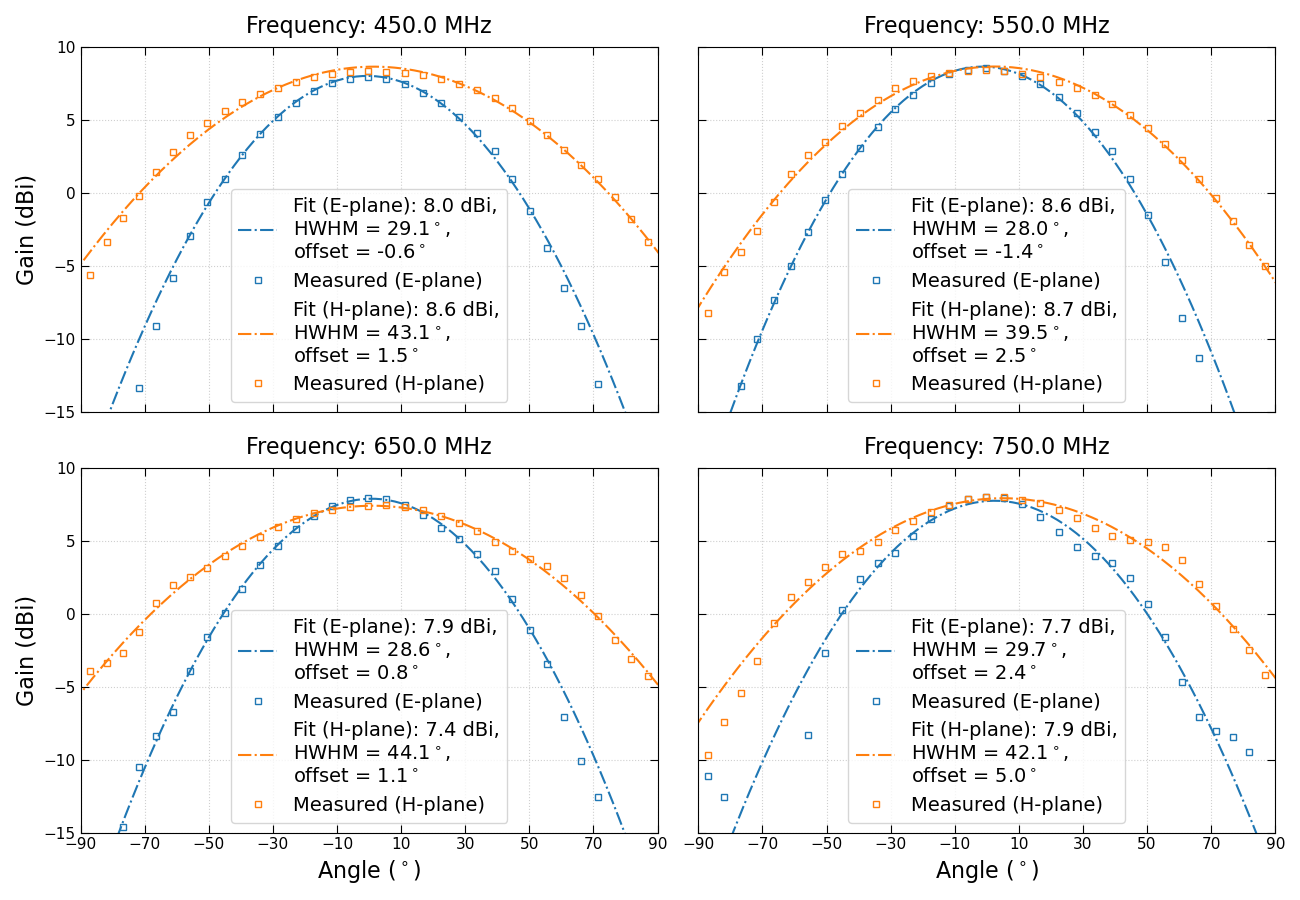}

    \caption{
        Similar as Figure \ref{fig:RadPattern-MM}, the measured E-plane (orange curves) and H-plane (blue curves) radiation pattern of ASIAA LPDA at 450, 550, 650, 750 \si{MHz} calibrated in RF anechoic chamber, and the corresponding Gaussian fitting around the main lobe.        
    }

    \label{fig:RadPattern-ASIAA}
\end{figure}

\begin{figure}[htb!]
    \centering

        \includegraphics[width=0.75\textwidth ]{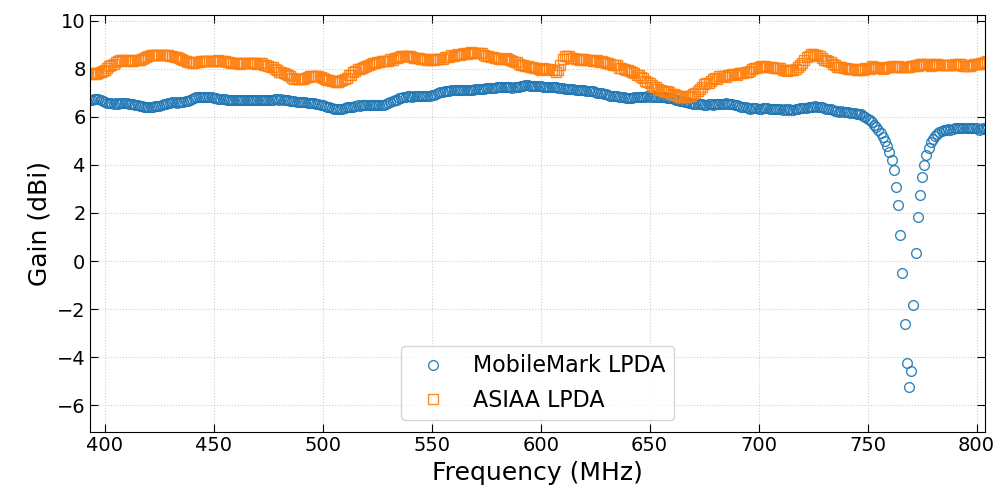}

    \caption{
        The measured gain of commercial (blue) and ASIAA (orange) LPDA calibrated in RF anechoic chamber. See Appendix \ref{sec:AntCal} for calibration details.
    }

    \label{fig:GainBoresight}
\end{figure}

The gain and radiation patterns of both types of LPDAs were measured in a RF anechoic chamber, the details of which are described in Appendix \ref{sec:AntCal}.
The measured E-plane and H-plane radiation patterns of both antenna types are shown in Figure~\ref{fig:RadPattern-MM} and \ref{fig:RadPattern-ASIAA}, and the gain response at boresight is shown in Figure \ref{fig:GainBoresight}.
The commercial LPDA exhibits a realized gain of $\approx$\SI{6}{\dBi} and across the \SIrange{400}{800}{\MHz} frequency band, with a gain drop near \SI{770}{\MHz}, in contrast to the nominal \SI{9}{\dBi} in the manufacturer specifications. 
The full width at half maximum (FWHM) of the antenna main lobe is $\approx$\ang{60} in the E-plane and $\approx$\ang{120} in the H-plane, both of which are well approximated by Gaussian profiles. 
The forward radiation pattern of each LPDA is approximated by a 2D-Gaussian profiles for sensitivity calibration in Sec.~\ref{sec:cal}.


The custom ASIAA LPDA was developed in parallel for superior performance compared to the commercial one. It is designed to have a wider frequency coverage at \SIrange{300}{800}{\MHz} with a higher gain of $\approx$\SI{8}{\dBi} to reduce ground noise pickup and shadowing by adjacent antennas. A key feature of its design is the direct integration of the LNA into the antenna feed, to reduce the receiver noise temperature. In addition, the custom design incorporates improved weatherproofing, as the performance of some of the commercial units was found to degrade in the humid environment of the Fushan site.
The LPDA design is detailed in \cite{Chiong2025}.
The LPDA design follows the procedure outlined in \citet{Balanis2016}. 
The geometric spacing factor is reduced to half the value suggested in the reference to compact the height of the LPDA from \SI{1725}{\mm} to \SI{950}{\mm} for easier transport and deployment. The trade-off of reducing the spacing factor is a poorer return loss at higher frequencies.
 %
 The antenna gain at boresight is \SI{7.9 \pm 0.6}{dBi} across the \SIrange{300}{800}{\MHz} band. The FWHM beam widths are \ang{60} in the E-plane and \ang{90} in the H-plane \citep{Chiong2025}, as shown in Figure \ref{fig:RadPattern-ASIAA}.
The commercial antennas have been phased out and fully replaced by the custom ASIAA LPDAs in March 2026.
 
\subsection{RF Front-end}
\label{sec:FEM}

\begin{figure}
    \centering
    \includegraphics[width=0.6\textwidth]{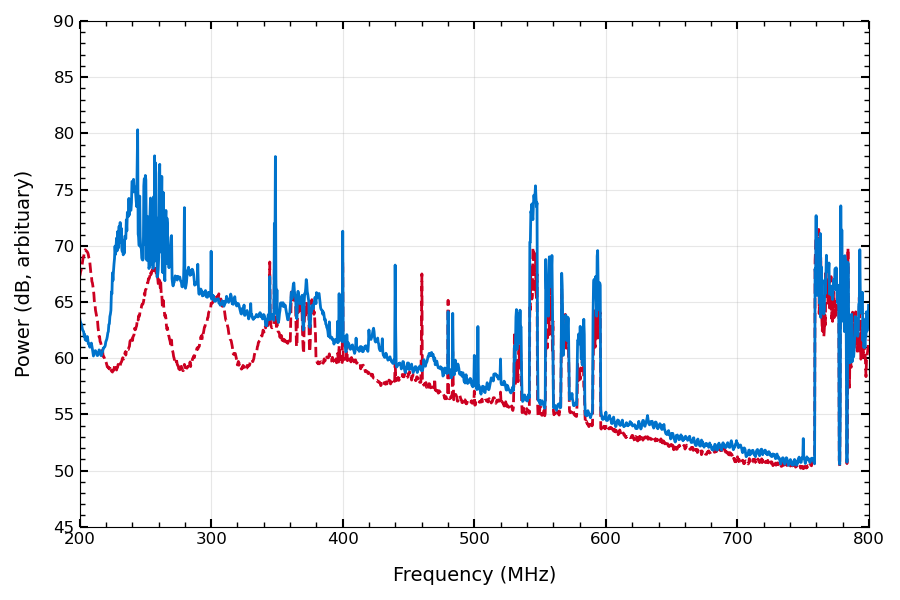}
	\caption{
        Example of power spectra taken with   commercial LPDA (red dashed curve), which has frequency cutoff at $\sim$\SI{350}{\MHz} versus with ASIAA LPDA (blue solid, without high-pass filter), directly read out from the RFSoC board prior to beamforming and bit truncation.
        The slope results mainly from the response of amplifiers and long coaxial cables.
        Besides the RFI from DTV, 4G-LTE, and Satcom signals, as shown in Figure~\ref{fig:RFsurvey}, there is moderate broadband RFI from the outdoor cabinets.
	}
\label{fig:RxResponse}
\end{figure}

There are two RF front-end configurations depending on the LPDA design.
The commercial LPDA is connected to the front-end module via a short coaxial cable, where the received signal is first amplified by an LNA inside the FEM. Mini-Circuits ZX60-P33ULN+ is selected as the LNA due to its low noise temperature of \SI{35}{\K}. 
A high-gain Mini-Circuits ZKL-1R5+ power amplifier is used as the second-stage amplifier to compensate for the insertion loss of the following \SI{40}{\m}-long low-loss CFD-400 coaxial cable. 
For ASIAA LPDA, the LNA is integrated with the antenna feed to lower the receiver noise temperature, and is then connected to the FEM housing other components, including an additional \SI{300}{\MHz} high-pass filter, with a \SI{2}{\m}-long coaxial cable (Figure~\ref{fig:front-end}). 
The RF components are housed within an RFI-shielded box inside a weatherproof ABS enclosure.
These two amplifiers are powered by direct current carried by the cable with bias tees at both ends.
A low-cost Nooelec LaNA amplifier is served as the third-stage amplifier before the signal goes to the RFSoC board, with a low-pass filter Mini-Circuits SLP-750+ with \SI{3}{\dB} cutoff frequency of \SI{780}{MHz} inserted at its input to prevent aliasing and saturation by 4G-LTE signal. 
16 of the antennas have an additional \SI{10}{\dB} attenuator pad inserted for experimenting different input power levels to the RFSoC board.
The length of the coaxial cables between the antenna assembly, the FEM, and the outdoor cabinets is \SI{2}{\m} and \SI{40}{\m}, respectively. 
The gain and noise budgets of the front-end are shown in Table \ref{tab:budgets}.
The overall front-end gain of $\sim\SI{75}{\dB}$ boosts the signal power to around \SI{-10}{\dBm} which is optimal for the RFSoC boards, while keeping the receiver noise temperature at around \SI{37}{\K}.
Examples of power spectra taken by individual BURSTT receivers with two different antennas are shown in Figure~\ref{fig:RxResponse}, where the negative slope of $\sim$\SI{15}{\dB} over \SI{400}{\MHz} results mainly from the response of amplifiers and long coaxial cables.
Despite that there is broadband RFI from the outdoor cabinets, the strength is still manageable and can be suppressed by subsequent beamforming toward the sky in the back-end system.

\begin{deluxetable*}{lrrrr}
    \tablecaption{Gain and noise budgets of the receiver front-end \label{tab:budgets}}
    \tablehead{\colhead{} & \colhead{Gain (dB)} & \colhead{Accum. Gain (dB)}& \colhead{Noise (K)} & \colhead{System Noise (K)} \label{chartable}}
\startdata
LNA & 23.3 & 23.3 & 35 & 35.0\\
\SI{2}{\m} cable & -0.5& 22.8 & 39& 35.2\\
2nd amplifier & 40.9 & 63.7 & 290 & 36.7\\
\SI{40}{\m} cable & -3.3 & 60.4 & 330 & 36.7\\
3rd amplifier & 25.0 & 85.4 & 58 & 36.7\\
\SI{10}{\dB} pad & -10.0 & 75.4 & 2610 & 36.7
\enddata
\end{deluxetable*}

\subsection{Back-End System}
\label{sec:backend}

The BURSTT-256 back-end architecture employs a hybrid processing pipeline distributed across 16 Xilinx ZCU216 FPGA-based RFSoC boards and five Intel Xeon high-performance computing servers. The following is a brief overview, with details described in \cite{BURSTT_BACKEND}.

The amplified signals from 16 antennas in the same E-W row are fed into an RFSoC board, with a total of 16 boards. 
On each RFSoC board, the analog voltage waveform is digitized by time-interleaved analog-to-digital converters (ADCs), achieving an aggregate sampling rate of \SI{1.6}{\GHz} at 14-bit resolution.
The signal is converted to the frequency domain and channelized into 2048 channels using a polyphase filter bank (PFB) followed by a fast Fourier transform (FFT) over every \SI{2.56}{\micro\second} time frame, expanding the data bit width to 18 bits.
The first stage of 16-antenna beamforming is performed directly on the RFSoC board, forming 16 beams along the E-W axis, which will be detailed in Sec.~\ref{sec:beamforming}.
Due to limited data transport bandwidth over the \SI{100}{Gbps} Ethernet link between each RFSoC board and the host server, only a \SI{400}{\MHz} sub-band --- currently spanning \SIrange{400}{800}{MHz}, with provision for \SIrange{300}{700}{\MHz} in future operations --- is retained from the full \SI{800}{\MHz} RF bandwidth.
Furthermore, the complex output baseband voltages are recorded in two selectable quantization modes: standard survey observations utilize continuous 4-bit complex quantization (4-bit real and 4-bit imaginary), whereas calibration observations record higher-precision 16-bit complex voltages but reduce the data rate by saving only one out of every four frames.
The resulting data have \SI{2.56}{\micro\second} time resolution and \SI{400}{\MHz} bandwidth with 1024 channels, i.e., \SI{0.390625}{\MHz} frequency resolution.

Data from RFSoC boards are streamed to and buffered on four servers, each processing \SI{100}{\MHz} frequency bandwidth and maintaining a $\approx$\SI{15}{\second} ring buffer. These servers then perform the second stage of 16-row beamforming along the N-S axis, the methodology of which is also described in Sec.~\ref{sec:beamforming}.
All RFSoC boards are synchronized to common \SI{10}{\MHz} and 1 PPS reference signals distributed by a GPS-disciplined rubidium frequency standard (Brandywine NFS-220 Plus).

For real-time FRB detection, the intensity of beamformed baseband data is computed and integrated over 400 frames on each server, i.e., to \SI{1.024}{\milli\second} time resolution, and up-channelized from 1024 to 16384 channels. The intensity data from all 256 beams are then streamed to another server in charge of the real-time search of FRBs with dispersed pulses. 
%
On the pulse-searching server, the intensity data is further processed for RFI mitigation before performing a search on dispersed pulses.
This process includes subtracting the average intensity of all beams (i.e., the incoherent beam),  a time-domain high-pass filter subtracting the moving average with a \SI{31}{\ms} time window for each beam's intensity (i.e., detrending), and normalizing the intensity to have unit variance. 
The real-time FRB-like signal search adopts the \texttt{bonsai} package, which is based on the tree dedispersion algorithm, developed and highly optimized by CHIME/FRB project (see \citealt{CHIME2018},
\citealt{Merryfield2023} for details), and each beam is individually searched by a \texttt{bonsai} routine.
If any beam exceeds the specified DM and SNR thresholds (single-beam trigger), additional trigger criteria are applied across all beams to form a station-level trigger, detailed in \cite{BURSTT_BACKEND}. An event identification routine then classifies the candidate as RFI, known pulsar, or FRB candidate. Upon validation, the trigger is distributed to all BURSTT stations to initiate recording of buffered baseband data for later source localization with VLBI.

\begin{figure}
    \centering
    \begin{subfigure}[c]{.5\textwidth}
    \includegraphics[width=\linewidth]{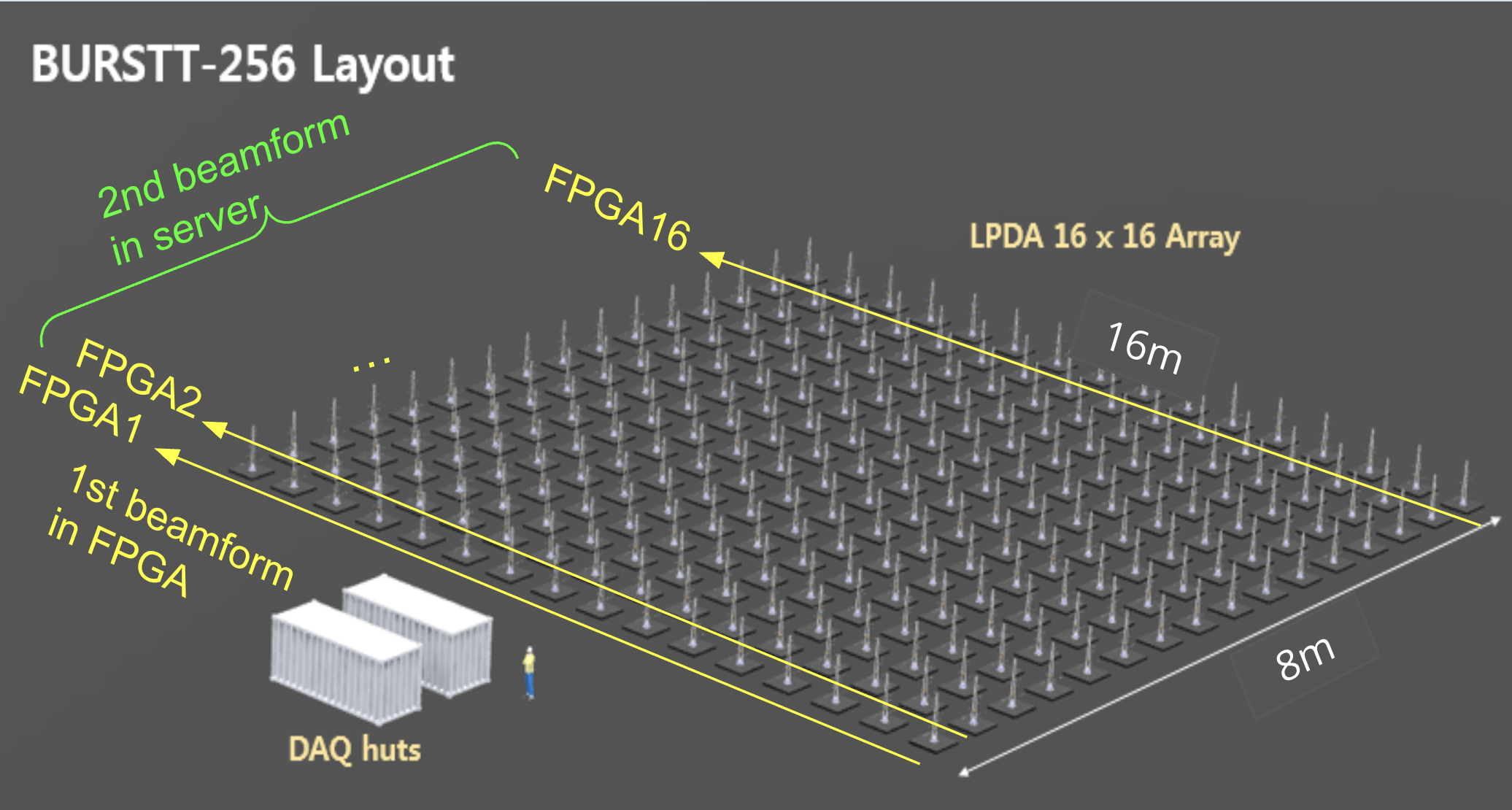}
    \end{subfigure}
    \hfill
    \begin{subfigure}[c]{.45\textwidth}
        \includegraphics[width=\linewidth]{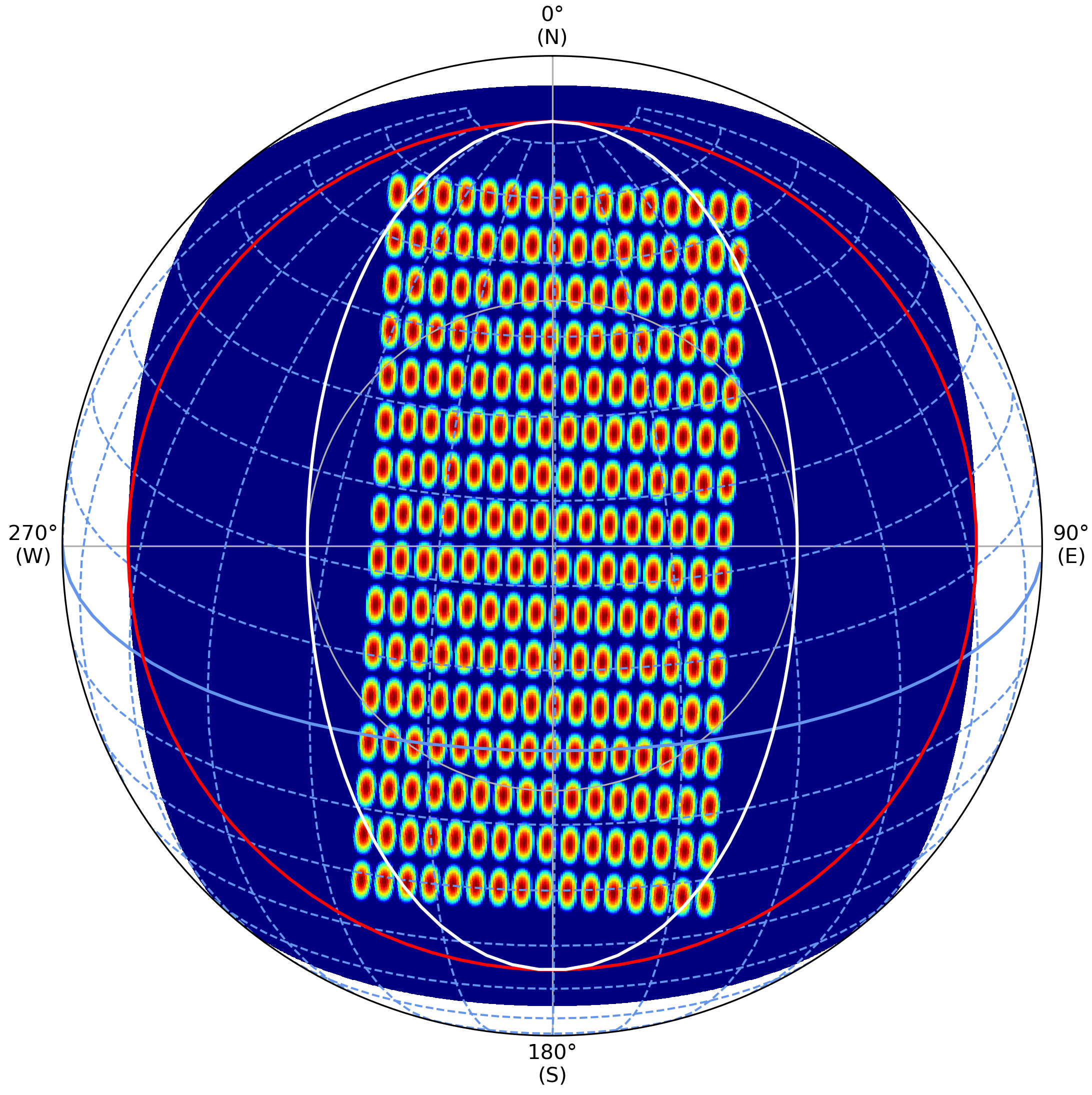}
    \end{subfigure}
    \caption{
    Left:  Schematics of the factorized beamforming process with two stages. In the first stage, signal from a row of 16 antennas along the E-W direction are combined within a single RFSoC board to synthesize 16 fan beams displaced along the same direction.
    In the second stage, data from the 16 separate RFSoC boards (along the N-S direction) are combined in the servers to further synthesize the final 2D pencil beams. 
    Right: Sky projection of the resulting 256 beams in horizontal coordinates. The thicker white ellipse denotes the half-power width of a single commercial LPDA. The red circle indicates a constant elevation of \ang{30}, approximating the local site horizon. Light blue dashed lines indicates the equatorial coordinate grid viewed from the site, with spacings of \ang{10} in Declination and \ang{15} in Right Ascension, where the celestial equator is highlighted in solid curve. Figure from \cite{BURSTT_BACKEND}.}
    \label{fig:beamform}
\end{figure}

\subsection{Beamforming Method}
\label{sec:beamforming}
The signal from each antenna is digitized individually and combined in the digital processing back-end system. 
The number of beams to be formed and searched is a free parameter limited by the processing power, and it is set to 256 in the initial operation.
For a wide FoV FRB survey, BURSTT-256 forms $16\times16$ stationary beams on the local sky, distributed over the LPDA's main lobe centered on the zenith. Figure~\ref{fig:beamform} shows a schematic of this process as well as the simulation of the resulting distribution of 256 beams in the sky. A full description of the back-end system is detailed in \cite{BURSTT_BACKEND}.

In short, the beamforming process is factorized into two steps. In the first beamforming step, the channelized baseband voltages from a linear array of 16 antennas along the E-W direction are represented by 18-bit integers on a single RFSoC board and synthesized into 16 beams through an 8-bit complex beamforming matrix (BFM). 
Due to the limit of the data bandwidth limit \SI{100}{Gbps} Ethernet, the beamformed data are truncated to 4-bit complex integers before being transferred to four servers. 
In the second step, the data from all 16 RFSoC boards are then synthesized into 16 beams along the N-S direction at the servers. 

The beamforming process can thus be represented by two $16 \times 16$ BFMs, $B_{X,mk}$ and $B_{Y,nj}$, implemented on FPGAs and servers, respectively, imposing on the complex voltage of individual antenna at frequency $\nu$ and time $t$, $V(\nu;t)$, yielding the synthesized beamformed output voltage $B(\theta_{X,m}, \theta_{Y,n},\nu;t)$ steered toward zenith angles $\theta_{X,m}$ along the E-W direction ($X$) and $\theta_{Y,n}$ along the N-S direction ($Y$), with $\theta \in [-90^\circ,90^\circ]$,

\begin{equation}
    B(\theta_{X,m}, \theta_{Y,n},\nu;t) = \sum_{k=0}^{N_{X}-1} B_{Y,nk}(\theta_{Y,n},\nu) \cdot \sum_{j=0}^{N_{Y}-1} B_{X,mj}(\theta_{X,m},\nu) V(x_j,y_k,\nu;t),
\end{equation}
where $j$ and $k$ denote the antenna position indices along the E-W and N-S axes, respectively,  for a $16\times16$ antenna array ($N_{X}=N_{Y}=16$). The beam indices $m$ and $n$ likewise range from 0 to 15.

Each of the BFM consists of three terms,
\begin{align}
    B_{X,mj}(\theta_{X,m},\nu) &= A_{X,mj}(\nu) \cdot \exp[i\Phi_{j}(\nu)] \cdot \exp[i\omega\tau_{\mathrm{gX},mj}(\nu)] , \\
    B_{Y,nk}(\theta_{Y,n},\nu) &= A_{Y,nk}(\nu) \cdot \exp[i\Phi_{k}(\nu)] \cdot \exp[i\omega\tau_{\mathrm{gY},nk}(\nu)]. 
    \label{eq:BFM}
\end{align}
The geometric delay term ($\tau_{\mathrm{g}}$) for each array element determines the direction in which a beam is synthesized. 
%
In practice, the amplitude scaling term ($A$) is needed for passband equalization and to accommodate the data within the 4-bit dynamic range, and the phase term $\Phi(\nu)=\omega\tau + \phi_{\mathrm{disp}}(\nu)$ is required to correct for the instrumental delay and dispersion of individual antennas  and RFSoC boards.
The passband equalization term and the instrumental delay of each receiver channel are calibrated on site using bright celestial radio sources with known flux density. The calibration procedure and the results will be detailed in Sec.~\ref{sec:cal}.

 At each beamforming stage, a set of geometric delay terms was chosen such that the resulting 16 beams are synthesized in an evenly spacing along each direction, with antenna 0 of each row in the west end and RFSoC board 0 at the south end as reference,
\begin{align}
    \tau_{gX,mj} &= \frac{jd_{X}}{c} \sin\theta_{X,m} = \frac{jd_{X}}{c}[ \frac{\lambda_0}{d_{X}}\frac{m+b_{0X}}{N_{X}}], \\
    \tau_{gY,nk} &= \frac{kd_{Y}}{c} \sin\theta_{Y,n} = \frac{kd_{Y}}{c}[ \frac{\lambda_0}{d_{Y}}\frac{n+b_{0Y}}{N_{Y}}],
    \label{eq:bf_angle}
\end{align}
where $d_{X}=\SI{1.0}{\m}$ and $d_{Y}=\SI{0.5}{\m}$ are the E-W and N-S antenna spacing, respectively, and the reference wavelength $\lambda_0$ is chosen to be that of \SI{400}{\MHz}. The beam offset $b_0$ is a free parameter to shift all beams toward a desired direction along each axis, currently set to $-7.5$ to center the beams around the zenith.
The resulting beams each have $\ang{2.5} \times \ang{5}$  beam width in the E-W and N-S directions, respectively, with a total instantaneous sky coverage of $\approx$\SI{3200}{\deg\squared}, distributed as shown on the right of Figure~\ref{fig:beamform}.
The upcoming upgrade will double the number of beams to 512 to reach the proposed coverage of $\approx$\SI{6000}{\deg\squared}.


\section{System Characterization and Performance}
\label{sec:cal}

To characterize and calibrate the receiver response, bright persistent radio sources with known position and radio flux density such as the Sun ($\sim$\SI{5e5}{\Jy} at \SI{400}{\MHz}), Cygnus A ($\sim$\SI{5000}{\Jy}), Cassiopeia A ($\sim$\SI{5000}{\Jy}), Taurus A ($\sim$\SI{1000}{\Jy}), are used as calibrators \citep{Perley2017}.

\subsection{Calibrating Instrumental Delay and Passband Equalization}
\label{sec:16bit-cal}

%
The Sun, the brightest celestial radio source as seen from Earth, is used to calibrate the frequency-dependent phase and gain responses of individual RF receiver channels. 
The receiver phase response can be decomposed into a linear term, corresponding to the instrumental delay, and a non-linear residual term associated with the dispersive effects. The gain response is calibrated to achieve proper passband equalization across frequency and receiver channels.
The calibration is performed in two steps: first among 16 analog receiver channels within a single RFSoC board for first-stage beamforming, and subsequently across 16 RFSoC boards for second-stage beamforming. 
The derived instrumental phase responses and gain scaling factors are then incorporated into the beamforming matrices (Equation~\ref{eq:BFM}).

Two cross-correlation based approaches have been tested to calibrate instrumental delay and give consistent results.
%
For every pair between 16 antennas, the baseband data of full 16-bit resolution (complex voltages with \SI{2.56}{\micro\second} per time frame and 1024 frequency channels) of the $i$-th and $j$-th antenna, $V_i(\nu,t)$ and  $V_j(\nu,t)$, are cross-correlated and averaged over time ($\sim5000$ time frames) to obtain the time-average visibility,
    \begin{align}
        C_{ij}(\nu) &= \left\langle \frac{  V_i(\nu,t) \cdot V_j^*(\nu,t)  } { | V_i(\nu,t) | \cdot | V_j(\nu,t) |} \right\rangle_t ,
    \label{eq:xcor}
    \end{align}
which is normalized and bound between $[-1,1]$, where $\langle \rangle_t$ denotes the time average. All $C_{ij}$ constitute a $16 \times 16$ cross-correlation (visibility) matrix.

\begin{figure}[htb!]
    \centering
    \begin{subfigure}{.49\textwidth}
        \includegraphics[width=\textwidth ]{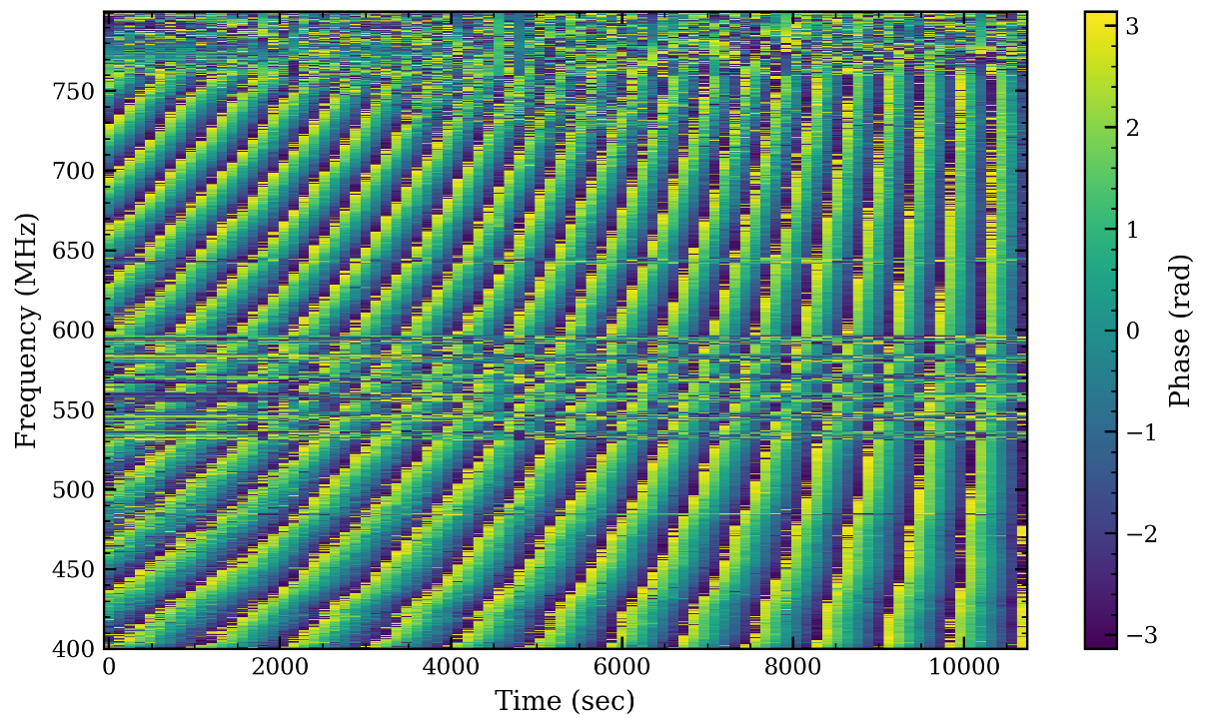}
    \end{subfigure}    
    \begin{subfigure}{.49\textwidth}
        \includegraphics[width=\textwidth ]{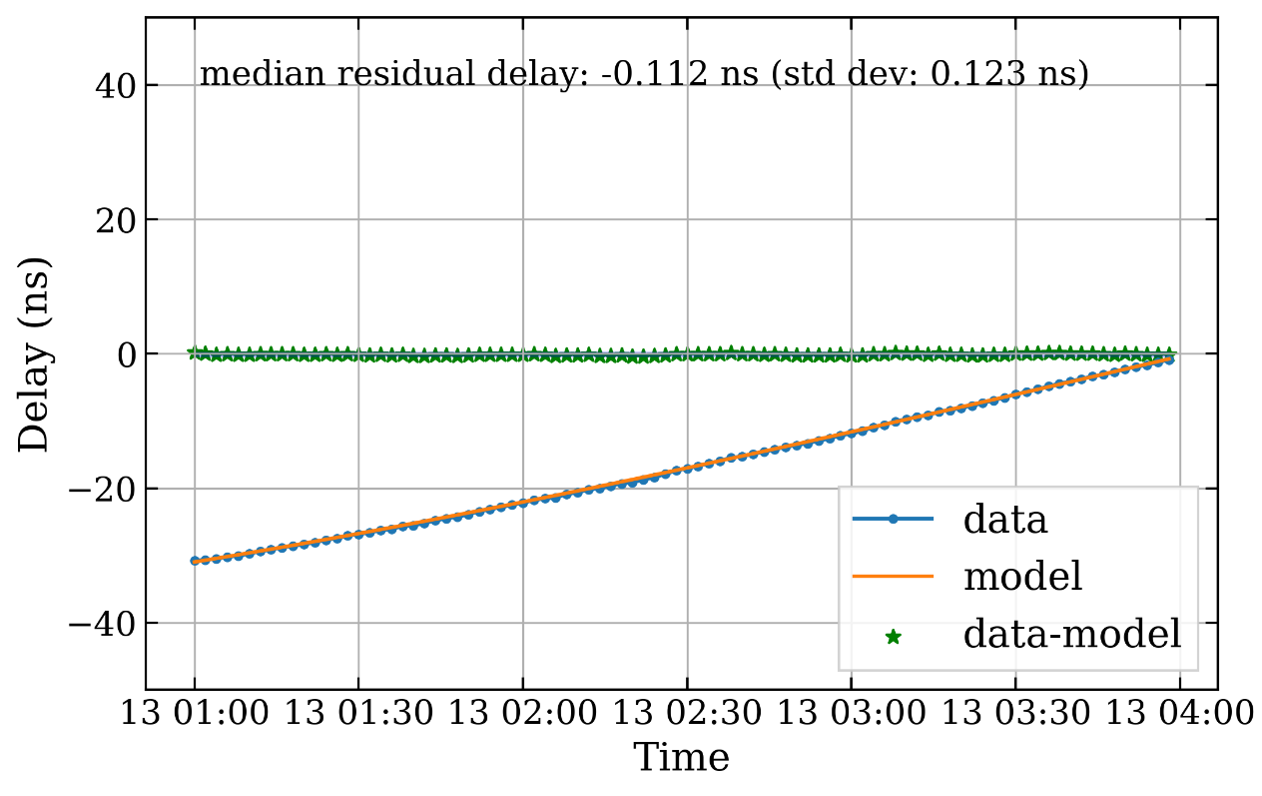}
    \end{subfigure}
    \caption{
        Example of calibrating individual receiver channels with the observation of solar interferometric fringe during 2024-01-13 09:00-12:00 local time. 
        Left: Cross-correlation phase as a function of frequency and time for the westernmost (0-th) and the second easternmost (14-th) antennas on the same N-S row with \SI{14}{\m} baseline. The average over time is subtracted to suppress persistent RFI. The fringe can be clearly seen.
        Right: The overall delay (blue circular markers) extracted from fitting linear phase variation versus frequency at each time sample.  After subtracting the expected geometric delay of the Sun (orange curve), the residual delay (green star-shape markers) is roughly constant over time, contributed mostly by the instrument delay, which is \SI{-0.112\pm0.123}{\ns} in the example.
    }
    \label{fig:SunFringe}
\end{figure}

\subsubsection{Interferometric Fringe Approach}

The first calibration approach observes the interferometric fringe of the Sun by taking the baseband data at a regular interval of a few minutes over 2-3 hours around noontime.
%
%
The cross-correlation matrix was then calculated for each antenna pair and each snapshot. 
The cross-correlation spectrum of each antenna pair over the entire observation period has its time average subtracted to suppress persistent RFI. The solar interferometric fringe can be observed in its dynamic phase spectrum. An example of the dynamic spectrum and the corresponding delay of an antenna pair is shown in Figure~\ref{fig:SunFringe}. 
Then, for each snapshot, the mean-subtracted cross-correlation phase as a function of the frequency between \SIrange{450}{760}{\MHz} is fitted linearly to extract the delay term by minimizing the squared sum of the residual phase.
%
The expected geometric delay of the Sun was computed based on the site location, the observation time, and the position of antenna pairs, and then it was subtracted from the extracted delay (e.g., orange curve in Figure~\ref{fig:SunFringe}). 
%
The orientation of the antenna array can also be left as a free parameter in the fitting and calibrated (Sec.~\ref{sec:array}).
The residual delay, which is nearly constant throughout the observation period, is assigned as the instrumental delay, as shown on the right panel in Figure~\ref{fig:SunFringe}.

\subsubsection{Spatial Filtering Approach}

The second approach takes only one snapshot when the Sun is in field of view instead of an extended observation, and hence it is more suitable for regular calibration. The gain scaling for passband equalization is also calibrated with this method. However, it requires a different method to extract the solar signal component from that of the ambient RFI, for which we followed the spatial filtering method described in \cite{Kocz2010}.

%
To perform spatial filtering, the covariance matrix $C_{ij}(\nu)$ in Equation~\ref{eq:xcor} is also computed for each frequency channel, and the $16\times16$ matrix is then eigen-decomposed into its constituent eigenmodes.
The core idea of this technique is that the covariance matrix captures the relative arrival times (phases) and powers of incoming signals across the array. In an ideal scenario containing multiple independent point sources and uncorrelated system noise, the decomposition partitions the matrix into distinct signal and noise subspaces, where each eigenvector represents the unique arrival direction of a source, and its corresponding eigenvalue denotes the signal's amplitude. 
%

\begin{figure}[htb!]
    \centering
        \includegraphics[width=0.9\textwidth]{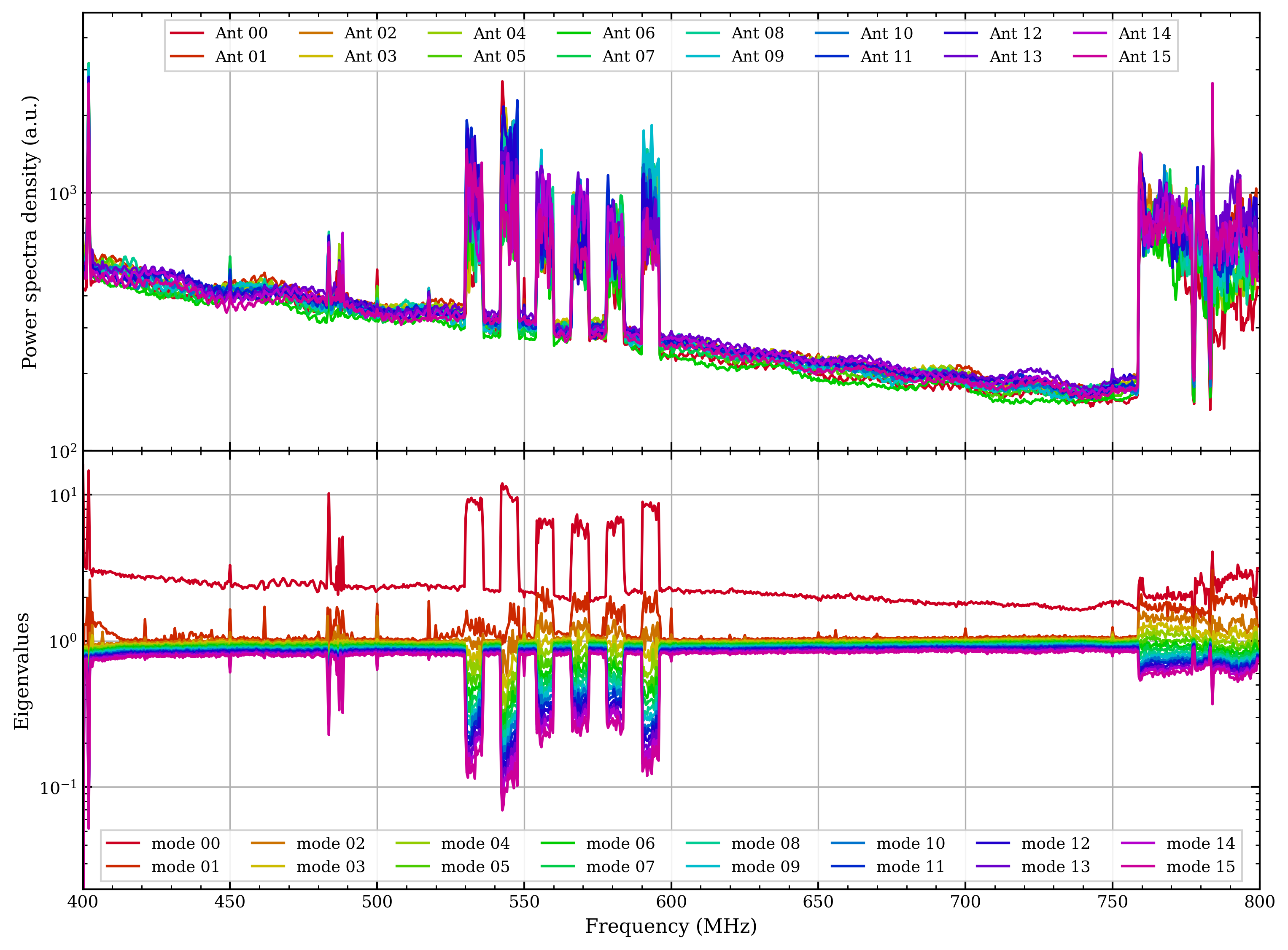}
    \caption{
        Top: the time-average power spectra recorded by 16 antennas connected to a single RFSoC board at Fushan station on Oct 17, 2025.
        Bottom: The corresponding noise spectra (eigenvalues) after eigen-decomposition of its covariance matrix. The leading mode with the highest strength has a continuum spectral component (brown color), which is expected to be predominantly from the solar signal. 
    }
    \label{fig:Eigen_Sun}
\end{figure}

An example of the eigen-decomposition result of our calibration observation is shown in Figure~\ref{fig:Eigen_Sun}. The leading eigenmode is expected to be dominated by the solar radio signal except at frequencies with narrowband RFI such as DTV and 4G-LTE. Therefore, by extracting only this principal eigenmode for calibration, we effectively zero out the other eigenmodes, which project out the interference and noise subspaces to mitigate the RFI.

Similarly to the previous approach, the expected geometric delay of the Sun is subtracted from the phase term of the leading eigenmode, and the antenna response is corrected for its amplitude term.
The leading eigenvector after subtraction is interpreted as the complex antenna solution, consisting of the bandpass and phase response, Equation~\ref{eq:BFM}.


The delay between RFSoC boards for the second-stage beamforming follows a similar calibration method with eigen-decomposition, based on the snapshot of 4-bit baseband data with first-stage beamforming, taken when the Sun transits through a beam close to the zenith. The resulting delay and normalization factors for each board are also provided to Equation~\ref{eq:BFM}.

Multiple calibrations of instrumental delay over several months show that the delay variation of a single receiver channel is $\sim$\SI{0.2}{\ns}, and the delays are consistent with the measured overall group delay of the receiver chain within $\sim$\SI{0.1}{\ns} in the lab with a vector network analyzer prior to deployment.

\subsection{System Equivalent Flux Density}
\label{sec:SEFD}

\subsubsection{Single-Antenna SEFD}
\label{sec:SEFD_Ant}

To verify that all 256 receiver signal chains exhibit comparable responses and to determine the relative amplitude weights required for optimal beamforming, we evaluate the per-element sensitivity across the frequency band through ad-hoc trials and after system changes.
 The sensitivity of the receiver as a function of frequency can be characterized by its system equivalent flux density (SEFD), defined as \citep{Thompson2017}:
\begin{equation}
    \mathrm{SEFD}(\nu) \equiv  \frac{ 2k_BT_{\mathrm{sys}(\nu)} }{ A_{\mathrm{eff}}(\theta,\phi,\nu)} = \frac{ 8 \pi k_B T_{\mathrm{sys}}(\nu) }{ G(\theta,\phi,\nu)\lambda^2 },
    \label{eq:SEFD}
\end{equation}
where $k_B$ is the Boltzmann constant, $T_{\mathrm{sys}}$ is the system noise temperature, $A_{\mathrm{eff}}$ is the antenna effective aperture, $G$ is the antenna gain (radiation pattern), $\lambda$ is the wavelength, and the leading factor of 2 comes from receiving unpolarized noise by a linearly polarized antenna.

Using the Sun as the primary calibrator, the SEFD of each antenna can be derived from the observed cross-correlation between antenna pairs.
In the absence of RFI --- which generally holds outside of the DTV and 4G-LTE bands --- the cross-correlation coefficient measures the coherent fraction of the signal, mainly from the Sun. The residual incoherent fraction consists of system noise, including receiver noise and antenna noise mainly consisting of the thermal emission from the ground and Galactic noise from the sky. Consequently, using the same 16-bit calibration data as in Section~\ref{sec:16bit-cal}, the baseline SEFD between antennas $i$ and $j$ can be determined from the measured cross-correlation coefficient as
\begin{align}
    |C_{ij}(\nu)| &\approx \frac{ S_{\mathrm{src}}(\nu) g(\theta,\phi, \nu) } {(\mathrm{SEFD}_{ij}(\nu)/ \sqrt{ \Delta\nu  \tau}) + S_{\mathrm{src}}(\nu) g(\theta,\phi, \nu) }, \\
    \mathrm{SEFD}_{ij}(\nu) &\approx (\frac{1- |C_{ij}(\nu)| }{ |C_{ij}(\nu)| }) S_{\mathrm{src}}(\nu) g(\theta,\phi, \nu)  \sqrt{ \Delta\nu  \tau}, 
\label{eq:SEFD2}
\end{align}
where the numerator is the received signal power and the denominator is the total received power. 
$S_{\mathrm{src}}(\nu)$ stands for the source flux density, and $g(\theta,\phi, \nu) \equiv G(\theta,\phi, \nu)/G_0$ represents the relative gain with respect to the boresight, to account for the beam attenuation factor toward the source. 
%
The main lobe of LPDA was approximated by a two-dimensional Gaussian beam, with the gain and beam width provided in Sec.~\ref{sec:antenna}. The square root term represents the radiometer equation's sensitivity improvement, given the frequency channel bandwidth $\Delta \nu =\SI{0.390625}{\MHz}$ and the integration time $\tau = \SI{2.56}{\micro\second}$.
%
As the baseline SEFD equals to the geometric mean of the individual antenna SEFDs, $\mathrm{SEFD}_{ij} = \sqrt{\mathrm{SEFD}_{i} \cdot \mathrm{SEFD}_{j}} $, the antenna SEFDs can be derived by taking the logarithm of the relation and solving the pseudo-inverse matrix.

The solar radio flux density, $S_{\mathrm{src}}(\nu)$, in the \SIrange{300}{800}{\MHz} range is typically of the order of \SI{0.5}{\mega\Jy}. 
For calibration, we adopt the \SI{410}{\MHz} and \SI{610}{\MHz} solar radio flux density recorded at Learmonth Solar Observatory \citep{Learmonth}\footnote{\url{https://www.sws.bom.gov.au/World_Data_Centre/1/10}}. It is one of four stations in the global Radio Solar Telescope Network (RSTN), which provides solar flux data with 1-second time resolution \citep{Guidice1981,NOAA_NCEI_RSTN}.
We found that the reported solar flux density has a $\approx20\%$ difference between RSTN stations; since the stations have similar instruments and the same calibration procedures, this indicates that the uncertainty of intensity reported by Learmonth should also be $\approx20\%$, which is consistent with a previous systematic study by \cite{Giersch2022}.
The flux density is linearly interpolated across the full frequency range of \SIrange{400}{800}{\MHz} as an approximation.

\begin{figure}[htb!]
    \centering
    \includegraphics[width=.8\textwidth ]{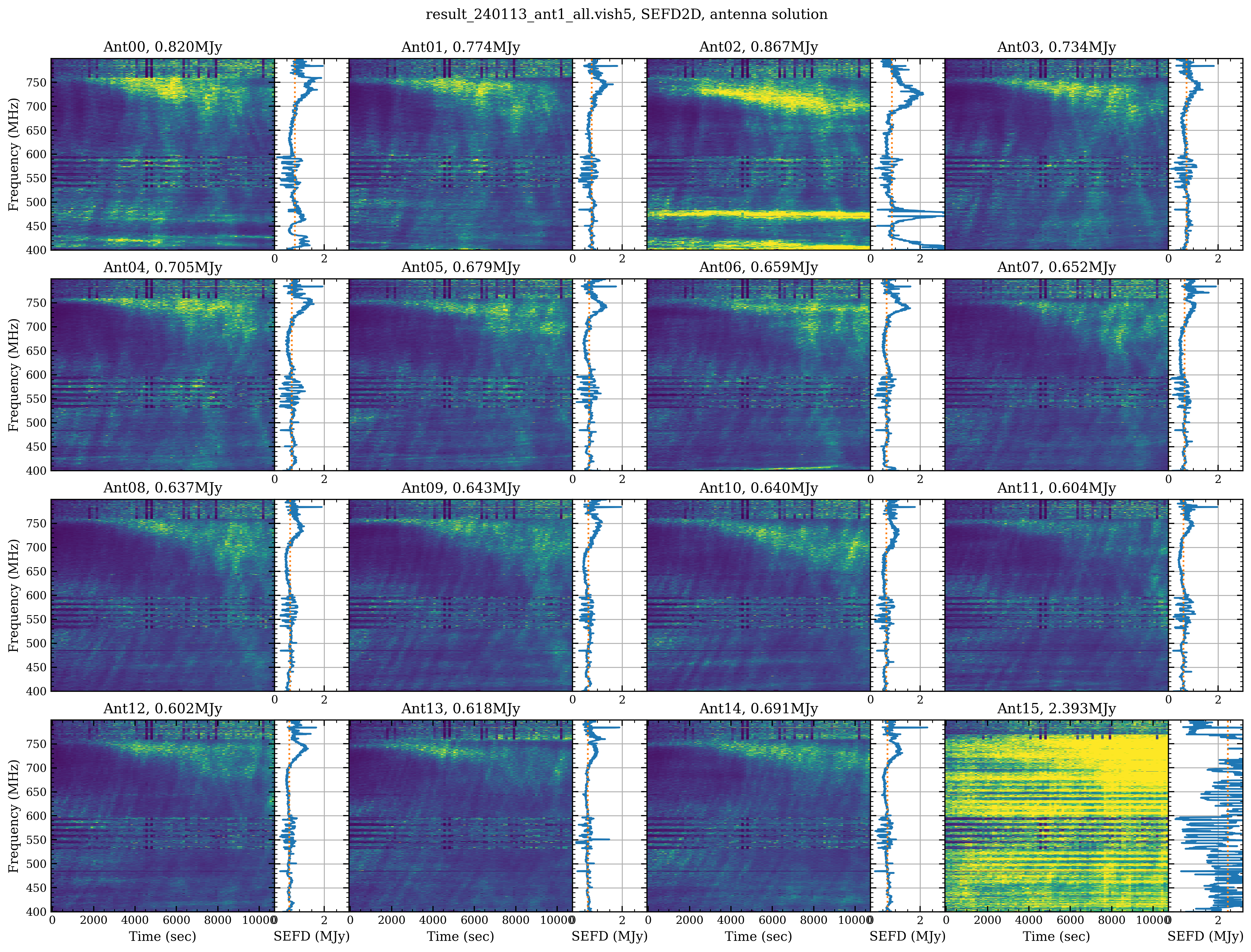}

    \caption{
        The derived SEFDs of the 16 individual receiver channels with commercial LPDAs as a function of frequency and time (left panel) and their median over time (right panel), from the solar interferometric fringe observation during 2024-01-13 09:00-12:00 local time (same as Figure~\ref{fig:SunFringe}). 
        The SEFD is scaled relative to \SI{400}{\MHz} to remove $\lambda^{-2}$ dependence for better visualization, where the the median value over frequency is shown on the title of each panel and indicated by orange dotted lines at right panel. In this example, the SEFD is between \SIrange{0.6}{0.8}{\mega\Jy} except the antenna 15 has an abnormally high value, which was thus excluded from the beamforming and later replaced.
    }
    \label{fig:SEFD_Sun}
\end{figure}

Using the method described above, the measured SEFD of the commercial LPDA is typically $\sim$\SI{1.5}{\mega\Jy} at \SI{400}{\MHz}, corresponding to $T_{\mathrm{sys}}\approx$\SI{100}{\kelvin}. 
Figure~\ref{fig:SEFD_Sun} illustrates the SEFD measurement results of 16 individual antennas in one E-W row on January 13, 2024, to demonstrate receiver calibration and monitoring. Most elements exhibit mutually consistent SEFD profiles, with median value throughout the period overall follows $\lambda^{-2}$ scaling except at DTV and 4G-LTE bands, while defective units exhibiting anomalous responses (e.g., due to antenna degradation) were flagged, excluded from the subsequent beamforming, and later replaced.
%
The nearly horizontal, brighter stripe features visible across every SEFD plot in Figure~\ref{fig:SEFD_Sun} likely result from the antenna coupling with adjacent ones, as these features persist across observing dates but systematically shift when the N-S antenna spacing was adjusted from \SI{1}{\m} to \SI{0.5}{\m}. 
Applying the same method, the ASIAA LPDAs yield an improved SEFD of $\sim$\SI{1.3}{\mega\Jy}, a gain-driven reduction attributed to a lower ground-noise pickup.

We note that while element-to-element consistency is high within any individual observation, the absolute SEFD varies between  \SIrange{0.5}{2.5}{\mega\Jy} across different observing dates, the causes of which have not been fully understood. This measured range nonetheless encompasses the expected theoretical value detailed in the next paragraph.
In any case, absolute radiometric precision is not critical at this calibration stage, as the primary objective is to determine the instrumental delays and relative amplitude weights for coherent beamforming.

For comparison, the expected single-antenna SEFD is estimated with the following steps. The antenna radiation pattern is approximated by 2D-Gaussian function with the antenna gain of \SI{6}{dBi} for commercial LPDA at the boresight (\SI{8}{dBi} for ASIAA LPDA). The receiver temperature of the front-end module is $\approx$\SI{40}{\kelvin}.
The expected antenna noise temperature at a given time, $T_{\mathrm{ant}}(t)$, equals the integral of the brightness distribution $B(\theta,\phi,\nu)$ convolved with the antenna effective area $A_{\mathrm{eff}}$, or equivalently to the weighted average of brightness temperature by the antenna gain pattern $G_{\mathrm{ant}}$ over all directions \citep{ARIANNA2017},
    \begin{align}
       T_{\mathrm{ant}}(t) 
       &= \frac{1}{2k_B} \int  B(\theta,\phi,\nu,t) A_{e} (\theta,\phi,\nu) d\Omega \nonumber \\ 
        &= \frac{1}{2k_B} \int  [\frac{2 k_{B} \nu^{2}  T_{B}(\theta,\phi,\nu,t)}{c^{2}}] [\frac{c^{2}G_{\mathrm{ant}}(\theta,\phi,\nu)}{4 \pi \nu^{2}}] d\Omega   \nonumber  \\
        &= \frac{1}{4\pi}\int  T_{B}(\theta,\phi,\nu,t)G_{\mathrm{ant}}(\theta,\phi,\nu) d\Omega 
    \label{eq:AntNoise}
    \end{align}
    where the factor of $1/2$ is the same polarization loss term as in Equation~\ref{eq:SEFD}.
    The brightness is approximated by the Rayleigh–Jeans law at radio frequencies and is related to the brightness temperature $T_{B}(\theta, \phi, f,t)$, which is equal to the sky noise temperature $T_{\mathrm{sky}}(\theta,\phi,t)$ for angles above the local horizon,  whereas a constant \SI{300}{\kelvin} ground temperature is assumed at below the horizon. The local terrain is taken into account based on photogrammetry taken in the field.

The sky noise is dominated by the Galactic noise at \SIrange{300}{800}{\MHz}, for which we adopt the map from LFmap \citep{LFmap}. 
After convolving with the assumed antenna radiation pattern, the Galactic noise contributes \SIrange {45}{65}{\kelvin} at \SI{400}{\MHz} depending on the local sidereal time, while the ground noise contributes to $\approx$\SI{10}{\kelvin}, mainly from the surrounding hills up to $\approx$\ang{30} elevation angle.
The overall expected system noise temperature at \SI{400}{\MHz} is therefore $\approx$\SIrange{95}{115}{\kelvin}, corresponding to an SEFD of \SIrange{1.6}{1.8}{\mega\Jy} at \SI{400}{\MHz}.

\subsubsection{SEFD After Beamforming}
\label{sec:SEFD_BF}


To measure the SEFD after beamforming (the array SEFD), fainter radio celestial calibrators are observed instead of the Sun, which would saturate the limited 4-bit dynamic range.
For each calibrator, the beamformed intensity data of \SI{1}{\milli\second} resolution of all beams were recorded continuously for about three hours when the source passed through the beams formed within the field of view. The resulting temporal profiles are compared with the expected ones from the simulations. 
Adopting the known radio flux intensity of Cygnus A  and Taurus A from \cite{Perley2017}, the measured signal-to-noise ratios (SNRs) yield the SEFDs of individual synthesized beams.

Applying the approach, the SEFD derived from Cygnus A and Taurus A is $\approx$\SI{12}{\kilo\Jy} and $\approx$\SI{11}{\kilo\Jy}, respectively. These values are approximately a factor of 2 higher than the theoretical value of ideal beamforming, $\approx$\SI{6}{\kilo\Jy}, which scales inversely with the number of antennas, $N_{\mathrm{ant}}$.
In principle, the array SEFD is expected to improve beyond the $N_{\mathrm{ant}}^{-1}$ scaling because coherent beamforming narrows the primary reception pattern, thereby suppressing the ground spillover and terrestrial RFI entering through the sidelobes.

Futhermore, the $\approx 10\%$ difference between the sensitivity derived from the two calibrators has persisted over several months of monitoring. As the declination of Taurus A is \ang{+22;01;03}, which is nearly overhead at the main station, while that of Cygnus A is \ang{+40;44;02}, this discrepancy is likely stems from inaccurate approximation of the antenna radiation pattern with a 2D-Gaussian pattern, especially along the N-S axis where the antennas have a shorter spacing, which is expected to be altered in the presence of adjacent antennas.

%
The cause of sub-optimal array SEFD is under investigation and will be addressed in the upcoming paper \citep{BURSTT_overview}. Preliminary analysis suggests it is likely because the gain scaling factor for passband equalization is not optimal and leads to saturation of 4-bit dynamic range or over-weighting antennas that are less sensitive.
With replacement with ASIAA LPDAs and better choices of the scaling factors of data quantization and beamforming currently in progress, we expected the array SEFD to be improved to $\approx$\SI{4}{\kilo\Jy} in late 2026 (Table \ref{tab:summary}).

\subsection{Beamforming Verification}
\label{sec:BFCal}

\begin{figure}
\centering    \includegraphics[width=0.95\textwidth]{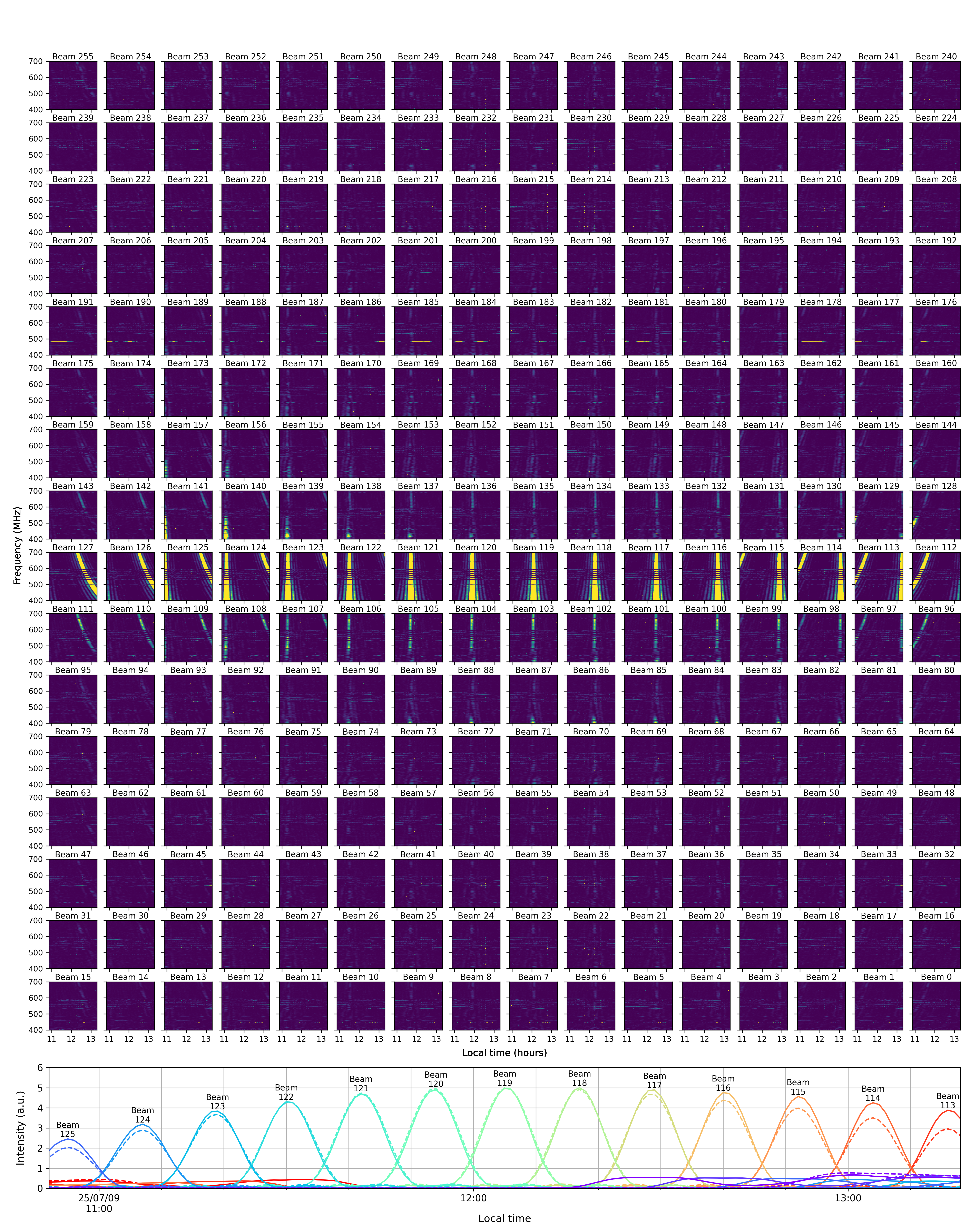}
	\caption{
        Intensity of the Sun after second-stage beamforming. The upper 16$\times$16 panels show frequency–time spectra around the solar transit in the 400--700~MHz band. On 9 July 2025, the Sun was located at (RA, Dec) = (07h13m09s, +22d22m17s), corresponding to a trajectory across row 7 (beams 112--127) in the 16$\times$16 beam layout centered on the zenith. The lower panel compares the observed (solid lines) and simulated (dashed lines) relative beamformed intensities of the Sun. The observed intensity is obtained by averaging the frequency-dependent intensity across 400--700~MHz shown in the upper panels. A total of 208 commercial LPDAs with frequency coverage from 400 to 800 MHz were used. 
        }
\label{fig:2ndBF-Sun}
\end{figure}

To verify the beamforming described in Sec.~\ref{sec:beamforming}, beamformed intensity data of all 256 beams were continuously recorded for about 2--3 hours when a calibrator source traversed the field of view, and their temporal variations are compared with those expected from the simulations. 
%
For example, the 256-beam dynamic spectra and the simulated and observed relative intensity of the Sun as a function of time after beamforming with 208 antennas (the rest 48 were under upgrade at the time) are shown in Figure~\ref{fig:2ndBF-Sun}.

The Sun transited through the expected beams and the timing is well consistent, indicating that the instrumental delay and the orientation of the antenna array are well calibrated.
The fading intensity at lower beams away from the zenith results from two effects. The antenna gain monotonically decreases with the angles from the boresight. The other is due to \ang{-3} offset of the antenna array orientation, which makes the resulting beams not aligned with a constant declination, and thus the Sun deviates more from the center of these beams when transiting through lower beams.

The magnitudes are agreed within $\sim30\%$ for beams at lower elevation angles.
The difference in magnitude likely results from the Gaussian antenna beam assumed in the simulation, which was derived from measurements of a single isolated antenna. Mutual coupling between adjacent antennas in the array  was therefore not taken into account. 
The coupling effect may become more significant at lower elevation angles and alter the radiation pattern.
The mutual coupling between LPDAs is currently being investigated through detailed array electromagnetic simulation. In addition, in situ characterization of the antenna response will be performed using a calibrated transmitter.
As BURSTT-256 focuses on discovering FRBs, the discrepancy is tolerable at the moment. Overall, the beamforming performs as expected.

\section{Conclusion}
\label{sec:conclusion}


The BURSTT project aims to conduct a comprehensive, high-cadence survey of FRBs in the local Universe. In its initial phase, we have successfully deployed the BURSTT-256 main station at the Fushan Botanical Garden in northern Taiwan. The multi-beam phased array initially utilized 256  commercial LPDAs operating at \SIrange{400}{800}{\MHz} to expedite system development and signal processing validation. 
Individual receiver gain and phase responses and real-time beamforming are calibrated using bright radio celestial sources.
The resulting beamforming has a consistent directional distribution with the simulation, and the array currently has an SEFD of $\approx$\SI{11}{\kilo\Jy} and an FoV of $\approx$\SI{3200}{\deg\squared}.
However, the array SEFD is currently higher than expected, likely due to sub-optimal choice of gain scaling factors.
The antennas have been progressively upgraded with custom-designed ASIAA LPDAs, which yield a higher overall gain operating at \SIrange{300}{700}{\MHz}.

To further enhance the discovery potential and sensitivity of the BURSTT main station, several critical upgrades are currently underway:

\begin{itemize}
    \item \textbf{Antenna and Frequency Optimization:} Commercial antennas have been fully replaced by the custom ASIAA LPDAs in March 2026, which high gain leads to a lower the system noise by reducing the pickup of the ground noise. The primary operational frequency band is shifted downward to \SIrange{300}{700}{\MHz}. This adjustment not only increases the  effective aperture due to longer wavelengths, but also avoid RFI originating from 4G-LTE signals in the \SIrange{750}{800}{\MHz} band.

    \item \textbf{Improved Sensitivity by Backend Optimization:} Reduced sensitivity after beamforming in initial commissioning was found to be related to the beamforming parameters, such as the normalization factor between antennas. The optimization is in progress to recover the sensitivity to the expected value, where an initial test suggests that an array SEFD of $\approx$\SI{4}{\kilo\Jy} is achievable.
    Other calibration techniques --- such as interferometric imaging on bright calibrator sources introduced by \cite{BURSTT_OGA16}, and the statistical method based on pulse detection rates of standard pulsars with known fluence distributions  (e.g., single pulses of PSR~B0329+54 and giant radio pulses of  PSR~B0531+21) --- will be conducted to independently verify the array's sensitivity. 
    
    \item \textbf{Expanded Sky Coverage:} The digital backend processing algorithm will be upgraded to double the number of searched beams from 256 to 512, which will nearly double the instantaneous sky coverage of the array, achieving a FoV of $\approx$\SI{6000}{\deg\squared}. 
    
    \item \textbf{Polarimetry Integration:} To enable FRB polarimetry, a co-located sub-array of 16 orthogonally polarized ASIAA LPDAs, oriented along the North-South axis, will be integrated into the processing pipeline.

    \item \textbf{RFI Mitigation via RFoF:} To reduce self-generated RFI, Radio Frequency over Fiber (RFoF) technology will be implemented to relocate the RFSoC boards and associated digital electronics away from the receiver antenna array.
\end{itemize}

These near-term developments lay the fundamental groundwork for the project's next phase, which will expand the main array to 2048 or more antennas. Ultimately, this expansion will enable routine, sub-arcsecond localization of FRBs, opening new avenues for identifying FRB host environments and multi-wavelength counterparts.

\section*{Acknowledgement}
BURSTT is supported by  the National Science and
Technology Council (NSTC), Taiwan, under grant No.~NSTC 113-
2923-M-001-003-MY3 and Academia Sinica, Taiwan, under a
Grand Challenge Seed fund under grant No.~AS-GCS-114-M02.
TH acknowledges support from the NSTC through grants 113-2112-M-005-009-MY3, 114-2123-M-001-002-, and 111-2112-M-005-018-MY3.
The authors also gratefully acknowledge the Taiwan Space Agency (TASA) in Taiwan for providing access to the RF anechoic chamber facilities and technical assistance during the antenna calibration measurements.

\appendix
\section{Measurement of Antenna Radiation Pattern}
\label{sec:AntCal}

To characterize different antennas deployed across the BURSTT stations, their radiation patterns in both the $E$-plane (the plane containing the dipole elements) and the $H$-plane (the orthogonal plane containing boresight) were measured in the RF anechoic chamber at the Taiwan Space Agency (TASA), which has inner dimensions of $13 \times 10 \times 8$\si{\m}. Three distinct antenna designs were evaluated: the commerci al Mobile Mark LPDA, a custom-designed ASIAA LPDA, and a custom-designed magneto-electric dipole (MED) antenna utilized at some of BURSTT outrigger stations \citep{Su2024,BURSTT_overview}.

\begin{figure}[ht!]

    \centering
    \includegraphics[width=0.6\textwidth]{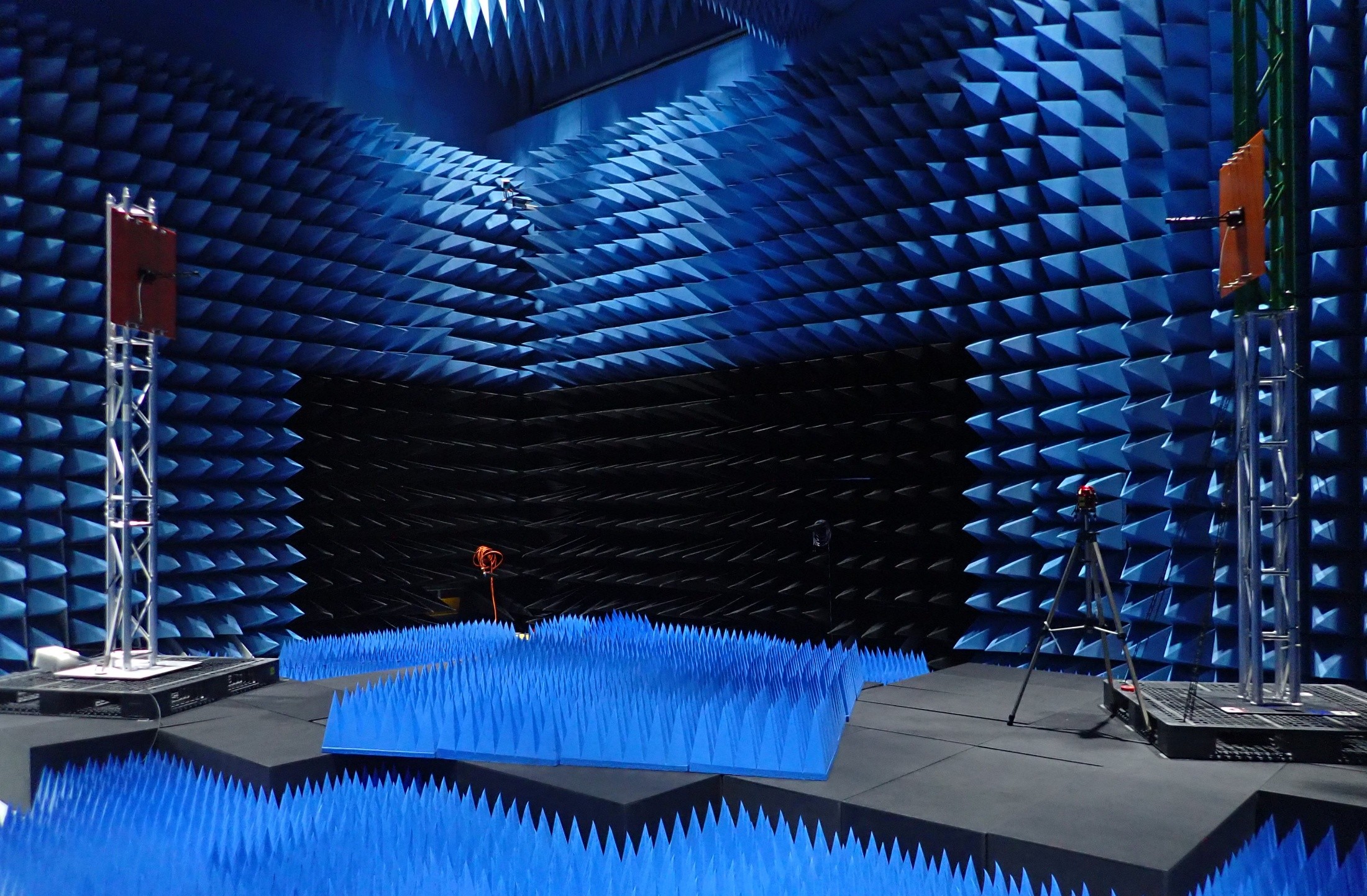}
    \caption{The measurement setup of antenna radiation pattern in RF anechoic chamber at TASA. The Rx antenna was installed on a turntable at left side, and the fixed Tx antenna was at the right, both of which were the commercial LPDAs in horizontal polarization for E-plane measurement in this case. The laser level meter beside the Tx antenna was removed after antenna alignment was done.
    \label{fig:AntCal}}
\end{figure}

Antenna gains and reception patterns were measured using a two-antenna transmission setup, as shown in Figure~\ref{fig:AntCal}. The transmitter (Tx) and receiver (Rx) antennas were separated by \SI{6.4}{\meter} along the $x$-axis, satisfying the far-field criterion, $R \ge 2D^2/\lambda$, where $D \approx \SI{0.5}{\meter}$ represents the maximum antenna dimension and $\lambda$ is the operational wavelength. 
Both the Tx and Rx antennas were mounted on an aluminum truss at a height of \SI{3.2}{\meter} above the chamber floor. The Tx antenna remained fixed, whereas the Rx antenna was mounted on a turntable defined as the coordinate origin, enabling azimuth rotation with an angular readout precision of $\approx\ang{1}$.
Measurements were performed with a vector network analyzer (VNA; Keysight N5222B) sweeping across the \SIrange{300}{1000}{\MHz} band, connected to the antennas inside the chamber through calibrated coaxial cables. 
The $E$-plane radiation pattern was characterized by aligning both antennas in horizontal polarization (Hpol) and rotating the turntable through an azimuth range of \ang{-90} to \ang{+90}. Similarly, the $H$-plane pattern was obtained by rotating both antennas to vertical polarization (Vpol) over the same angular span.
The realized gain profile is derived from the measured forward power transmission coefficient ($\lvert S_{21}\rvert^2$) using the Friis transmission formula \citep{Balanis2016}:
\begin{align}
    \lvert S_{21}\rvert^2 & \equiv \frac{P_{\mathrm{Rx}}}{P_{\mathrm{Tx}}} = G_{t}(\theta_{t},\phi_{t},\nu) (\frac{\lambda}{4\pi R} )^2 G_{r}(\theta_{r},\phi_{r},\nu),
    \label{eq:Friis}
\end{align}
%
where subscripts $t$ and $r$ denote the Tx and Rx antennas, respectively, and $G(\theta, \phi, \nu)$ is the angle-dependent gain. Polarization mismatch loss between the Tx and Rx antennas is negligible in this co-polarized configuration.

The transmitter boresight gain was determined from face-on  ($\phi_t=\phi_r=\ang{0}, \theta_t = \theta_r = 90^\circ$) co-polarized measurement in Vpol, using the same type of antenna at the Tx and Rx side, assuming both antennas have identical gain at boresight,
\begin{align}
    G_{t}(90^\circ, 0^\circ, \nu) &= \frac{4\pi R  }{\lambda } \sqrt{\lvert S_{21,\mathrm{boresight}} \rvert^2 },  
    \label{eq:Tx},
\end{align}
which was applied to the commercial LPDA and the MED antenna. For the ASIAA LPDA, an identical unit was unavailable during the test campaign; hence, the calibrated MED antenna was used as the reference transmitter.
%
The antennas were aligned to within \SI{3}{\cm} with a laser level meter, while the antenna separation $R$ was measured using a laser rangefinder with a precision of \SI{3}{\mm}. 
Although rotating the receiver turntable introduces a slight angular deflection of a few degrees  relative to the Tx boresight axis, the Tx boresight gain was applied to all other Rx scan angles as an approximation, since the FWHM beam width of the Tx is greater than \ang{60} for both type of antennas and the gain only varies slowly within the main lobe.

The radiation pattern of the Rx antenna is then determined by
\begin{align}
        G_{r}(\theta_{r},\phi_{r},\nu) &\approx  (\frac{4\pi R  }{\lambda } )^2 \frac{ \lvert S_{21}(\theta_{r},\phi_{r},\nu)\rvert^2 }{G_{t}(90^\circ, 0^\circ, \nu)}.
\end{align}
%
The true reception angle $\phi_r$ and the distance $R$  was computed by applying a geometric correction to the nominal azimuth readings from the turntable, to account for the offset of the Rx phase center relative to the rotation axis. 
%
Since the chamber is primarily designed for satellite testing above \SI{1}{\GHz}, its internal RF absorbers are known to be less effective at lower frequencies, exhibiting $\approx$\SI{20}{\dB} less absorption across \SIrange{300}{400}{\MHz} for those on the wall and even poorer absorption for those on the floor. 
This expected performance limitation is seen by our measurements, where non-negligible reflection introduced multipath interference, manifested as a frequency-dependent ripple in the measured $\lvert S_{21}\rvert^2$ and a distinct discrepancy between the Hpol and Vpol boresight responses at the lower band edge. Therefore, the results below \SI{400}{\MHz} were disregarded.
The resulting measured radiation patterns of commercial and ASIAA LPDAs are shown in Figure \ref{fig:RadPattern-MM}, \ref{fig:RadPattern-ASIAA}, and their frequency-dependent boresight gains are shown in Figure \ref{fig:GainBoresight} in Sec.~\ref{sec:Ant}.

\bibliographystyle{aasjournalv7}
\bibliography{reference}

\end{document}